# Solution of the Skyrme-Hartree-Fock-Bogolyubov equations in the Cartesian deformed harmonic-oscillator basis. (IV) HFODD (v2.07f): a new version of the program.

J. Dobaczewski[a,b,c,d,1] and P. Olbratowski[a,b,2]

[a] *Institute of Theoretical Physics, Warsaw University ul. Hoża 69, PL-00681 Warsaw, Poland*
[b] *Institut de Recherches Subatomiques, CNRS-IN$_2$P$_3$/Université Louis Pasteur, F-67037 Strasbourg Cedex 2, France*
[c] *Department of Physics and Astronomy, The University of Tennessee, Knoxville, Tennessee 37996, USA*
[d] *Physics Division, Oak Ridge National Laboratory, P.O. Box 2008, Oak Ridge, Tennessee 37831, USA*

**Abstract**

We describe the new version (v2.07f) of the code HFODD which solves the nuclear Skyrme-Hartree-Fock or Skyrme-Hartree-Fock-Bogolyubov problem by using the Cartesian deformed harmonic-oscillator basis. In the new version, all symmetries can be broken, which allows for calculations with angular frequency and angular momentum tilted with respect to the mass distribution. The new version contains an interface to the LAPACK subroutine ZHPEVX.

## NEW VERSION PROGRAM SUMMARY

*Title of the program:* HFODD (v2.07f)

*Catalogue number:*

*Program obtainable from:* CPC Program Library, Queen's University of Belfast, N. Ireland (see application form in this issue)

*Reference in CPC for earlier version of program:* J. Dobaczewski and J. Dudek, Comput. Phys. Commun. **131** (2000) 164 (v1.75r).

*Catalogue number of previous version:* ADML

*Licensing provisions:* none

*Does the new version supersede the previous one:* yes

*Computers on which the program has been tested:* SG Power Challenge L, Pentium-II, Pentium-III, AMD-Athlon

[1] E-mail: jacek.dobaczewski@fuw.edu.pl
[2] E-mail: przemyslaw.olbratowski@fuw.edu.pl

*Operating systems:* UNIX, LINUX, Windows-2000

*Programming language used:* FORTRAN-77 and FORTRAN-90

*Memory required to execute with typical data:* 10 Mwords

*No. of bits in a word:* The code is written in single-precision for the use on a 64-bit processor. The compiler option `-r8` or `+autodblpad` (or equivalent) has to be used to promote all real and complex single-precision floating-point items to double precision when the code is used on a 32-bit machine.

*Has the code been vectorised?:* Yes

*No. of lines in distributed program:* 43 308 (of which 19 386 are comments and separators)

*Nature of physical problem*
The nuclear mean-field and an analysis of its symmetries in realistic cases are the main ingredients of a description of nuclear states. Within the Local Density Approximation, or for a zero-range velocity-dependent Skyrme interaction, the nuclear mean-field is local and velocity dependent. The locality allows for an effective and fast solution of the self-consistent Hartree-Fock equations, even for heavy nuclei, and for various nucleonic ($n$-particle $n$-hole) configurations, deformations, excitation energies, or angular momenta. Similar Local Density Approximation in the particle-particle channel, which is equivalent to using a zero-range interaction, allows for a simple implementation of pairing effects within the Hartree-Fock-Bogolyubov method.

*Method of solution*
The program uses the Cartesian harmonic oscillator basis to expand single-particle or single-quasiparticle wave functions of neutrons and protons interacting by means of the Skyrme effective interaction and zero-range pairing interaction. The expansion coefficients are determined by the iterative diagonalization of the mean field Hamiltonians or Routhians which depend non-linearly on the local neutron and proton densities. Suitable constraints are used to obtain states corresponding to a given configuration, deformation or angular momentum. The method of solution has been presented in: J. Dobaczewski and J. Dudek, Comput. Phys. Commun. **102** (1997) 166.

*Summary of revisions*

1. Two insignificant errors have been corrected.
2. Breaking of all the three plane-reflection symmetries has been implemented.
3. Breaking of all the three time-reversal×plane-reflection symmetries has been implemented.

4. Conservation of parity with simultaneously broken simplex has been implemented.
5. Tilted-axis cranking has been implemented.
6. Cranking with isovector angular frequency has been implemented.
7. Quadratic constraint on tilted angular momentum has been added.
8. Constraint on the vector product of angular frequency and angular momentum has been added.
9. Calculation of surface multipole moments has been added.
10. Constraints on surface multipole moments have been added.
11. Calculation of magnetic moments has been added.
12. Calculation of multipole and surface multipole moments in the center-of-mass reference frame has been added.
13. Calculation of multipole, surface multipole, and magnetic moments in the principal-axes (intrinsic) reference frame has been added.
14. Calculation of angular momenta in the center-of-mass and principal-axes reference frames has been added.
15. New single-particle observables for a diabatic blocking have been added.
16. Solution of the Hartree-Fock-Bogolyubov equations has been implemented.
17. Non-standard spin-orbit energy density has been implemented.
18. Non-standard center-of-mass corrections have been implemented.
19. Definition of the time-odd terms through the Landau parameters has been implemented.
20. Definition of Skyrme forces taken from the literature now includes the force parameters as well as the value of the nucleon mass and the treatment of tensor, spin-orbit, and center-of-mass terms specific to the given force.
21. Interface to the LAPACK subroutine ZHPEVX has been implemented.
22. Computer memory management has been improved by implementing the memory-allocation features available within FORTRAN-90.

*Restrictions on the complexity of the problem*

The main restriction is the CPU time required for calculations of heavy deformed nuclei and for a given precision required. Pairing correlations are only included for even-even nuclei and conserved simplex symmetry.

*Typical running time*

One Hartree-Fock iteration for the superdeformed, rotating, parity conserving state of ${}^{152}_{66}\mathrm{Dy}_{86}$ takes about six seconds on the AMD-Athlon 1600+ processor. Starting from the Woods-Saxon wave functions, about fifty iterations are required to obtain the energy converged within the precision of about 0.1 keV. In case when every value of the angular velocity is converged separately, the complete superdeformed band with precisely determined dynamical moments $\mathcal{J}^{(2)}$ can be obtained within within forty minutes of CPU on the AMD-Athlon 1600+ processor. This time can be often reduced by a factor of three when a self-consistent solution for a given rotational frequency is used as a starting point for a neighboring rotational frequency.

*Unusual features of the program*

The user must have an access to the NAGLIB subroutine F02AXE, or to the LAPACK subroutines ZHPEV or ZHPEVX, which diagonalize complex hermitian matrices, or provide another subroutine which can perform such a task. The LAPACK subroutines ZHPEV and ZHPEVX can be obtained from the Netlib Repository at University of Tennessee, Knoxville: http://netlib2.cs.utk.edu/cgi-bin/netlibfiles.pl?filename=/lapack/complex16/zhpev.f
and
http://netlib2.cs.utk.edu/cgi-bin/netlibfiles.pl?filename=/lapack/complex16/zhpevx.f
respectively.

LONG WRITE-UP

# 1 Introduction

The method of solving the Hartree-Fock (HF) equations in the Cartesian harmonic oscillator (HO) basis was described in the previous publication, Ref. [1], which is below referred to as I. Two previous versions of the code HFODD, (v1.60r) and (1.75r), were published in Refs. [2] and [3], respectively, which are below referred to as II and III. The present paper is a long write-up of the new version (v2.07f) of the code HFODD. This extended version features the breaking of all the three plane-reflection symmetries, allows for the tilted-axis cranking, includes the solution of the Hartree-Fock-Bogolyubov (HFB) equations, and is fully compatible with all previous versions.

Information provided in I, II, and III remains valid, unless explicitly mentioned in the present long write-up.

In Section 2 we briefly review the modifications introduced in version (v2.07f) of the code HFODD. Section 3 lists all additional new input keywords and data values, introduced in version (v2.07f). The structure of the input data file remains the same as in the previous versions, see Section 3 of II. Similarly, all previously introduced keywords and data values retain their validity and meaning.

# 2 Modifications introduced in version (v2.07f)

## 2.1 Corrected errors

*2.1.1 Dimension of matrices used for the Gauss-Hermite integration.* Variable `NDGAUS`, which defines dimensions of matrices used for the Gauss-Hermite integration, was incorrectly linked to the maximum allowed number of the oscillator quanta, `NDOSCI`, and not to the maximum allowed numbers of the Gauss-Hermite nodes `NDXHRM`, `NDYHRM`, and `NDZHRM`. This could have caused problems for calculations that would use large number of nodes for small Harmonic Oscillator (HO) bases. This error was innocuous for the 'optimum' numbers of nodes, see Sec. II-3.5, and for any numbers of nodes smaller than optimum. Since using more nodes than optimum cannot increase precision of calculations, see Sec. I-4.3, there was no obvious motivation for the user to run the code in the mode that would have made the error harmful.

*2.1.2 Closing of the record file.* The record file, see Sec. II-3.9, was closed only after a given data set was completed, and not after each call to the RECORD subroutine. As a result, on some systems the record files left on the disc after a system crash, or when the run was killed, were unusable.

## 2.2 Breaking of all the three plane-reflection symmetries

The previous version of the code (v1.75r) assumed that one plane-reflection symmetry is conserved, i.e., that the $y$-simplex operator of Eq. (I-52) commutes with the single-particle (s.p.) Routhian. As a consequence, the angular-momentum vector was restricted to have only one non-zero component, the $y$ component (see Refs. [4, 5, 6] for a discussion of symmetries). Because of the recent interest in the so-called shears and chiral rotation phenomena [4], in the present version this restriction has been released. This allows for an arbitrary orientation of the angular momentum vector with respect to the mass distribution. It also allows for an arbitrary orientation of the angular frequency vector, i.e., for the so-called tilted-axis cranking. The extension is done at the expense of diagonalizing matrices that are twice larger than before, and by summing up the densities on a twice larger number of the Gauss-Hermite integration nodes. Since typical cases of the shears and chiral rotation do not involve parity-violating shapes, the problem can be simplified by employing the parity conservation, and such an option has also been implemented.

Altogether, one may classify the relevant symmetry conditions by considering the $y$-simplex ($\hat{S}_y$), $y$-signature ($\hat{R}_y$), and parity ($\hat{P}$). Since the product of any two of them is equal to the third one, one has five different possibilities of the conserved symmetry groups, see Table 1. These five options are governed by three switches, `ISIMPY`, `ISIGNY`, and `IPARTY` (see Sec. 3.2), as given in Table 1.

Table 1: Primary set of conserved and nonconserved symmetries allowed in the code HFODD version (v2.07f).

| Option | Symmetries | | | Switches | | |
|---|---|---|---|---|---|---|
| | $\hat{S}_y$ | $\hat{R}_y$ | $\hat{P}$ | `ISIMPY` | `ISIGNY` | `IPARTY` |
| P1 | conserved | conserved | conserved | 1 | 1 | 1 |
| P2 | conserved | nonconserved | nonconserved | 1 | 0 | 0 |
| P3 | nonconserved | conserved | nonconserved | 0 | 1 | 0 |
| P4 | nonconserved | nonconserved | conserved | 0 | 0 | 1 |
| P5 | nonconserved | nonconserved | nonconserved | 0 | 0 | 0 |

Switch `ISIGNY`, which was introduced in the previous versions, see Sec. II-3.3, is maintained in the way that ensures full compatibility of input data files. This is implemented in the following way. If switch `IPARTY` is set to $-1$ (the default value), than the code resets its value to `ISIGNY`, and stops in case switch `ISIMPY` is not set to 1 (the default value). In this mode, the present version (v2.07f) works exactly in the same way as the former versions. On the other hand, if switch `IPARTY` is set to 0 or 1, then the value of `ISIGNY` must be compatible with `ISIMPY` and `IPARTY` according to the group multiplication, Table 1, otherwise the code stops.

When either of the three symmetries $\hat{S}_y$, $\hat{R}_y$, or $\hat{P}$ is conserved, the single-particle Routhians acquire specific block-diagonal forms, cf. Ref. [6], and diagonalization of smaller matrices results in a faster execution time. In version (v2.07f), this is implemented for the conserved $y$-simplex and/or parity, but in case the $y$-signature alone is conserved it is not implemented yet. Hence, option P3 in Table 1 is enforced at the level of symmetries of densities [5], but in fact the code then operates as if no symmetry was conserved, and the execution time is not shorter. Similarly, within option P3, classification of single-particle states in terms of conserved $y$-signature is not available yet.

When the time-reversal and simplex or signature are conserved, the Kramers degeneracy allows for diagonalization of matrices only in one simplex or signature, respectively. This reduces the numerical effort by half. However, when the time-reversal alone is conserved, such a reduction is not possible, although the eigenstates do still obey, of course, the Kramers degeneracy.

In the present version, the antilinear symmetry $y$-simplex$^T$ [5], which is given by the operator $\hat{S}_y^T=\hat{T}\hat{S}_y$, can be either conserved or nonconserved. This is governed by switch ISIMTY=1 or 0, respectively, see Sec. 3.2. Of course, conservation of $y$-simplex$^T$ must by compatible with the conservation of the time-reversal ($\hat{T}$) and $y$-simplex ($\hat{S}_y$). The $y$-simplex$^T$ symmetry does not affect quantum numbers, but does affect shapes and directions of the angular momentum. In particular, its conservation confines the angular momentum to the $x$-$z$ plane. Similarly, conservation of $x$-simplex$^T$ and $z$-simplex$^T$, see Sec. II-3.3, which was already implemented in the previous versions, confines the angular momentum to the $y$-$z$ and $x$-$y$ plane, respectively. Therefore, the three antilinear symmetries allow for specific manipulations of the angular momentum vector, whenever the so-called planar rotation is required. Similarly, by conserving pairs of these antilinear symmetries one can restrict the angular momentum to one of the three Cartesian directions, and perform calculations with a conserved $x$-signature, $\hat{R}_x=(-1)^A\hat{S}_y^T\hat{S}_z^T$, or $z$-signature, $\hat{R}_z=(-1)^A\hat{S}_z^T\hat{S}_y^T$.

Table 2: Secondary set of conserved and nonconserved symmetries allowed in the code HFODD version (v2.07f).

| Option | Symmetries | | | Switches | | |
|---|---|---|---|---|---|---|
| | $\hat{S}_y$ | $\hat{T}$ | $\hat{S}_y^T$ | ISIMPY | ITIREV | ISIMTY |
| S1 | conserved | conserved | conserved | 1 | 1 | 1 |
| S2 | conserved | nonconserved | nonconserved | 1 | 0 | 0 |
| S3 | nonconserved | conserved | nonconserved | 0 | 1 | 0 |
| S4 | nonconserved | nonconserved | conserved | 0 | 0 | 1 |
| S5 | nonconserved | nonconserved | nonconserved | 0 | 0 | 0 |

In the previous versions, conserved $y$-simplex implied that the $y$-simplex$^T$ conservation was uniquely linked to the $\hat{T}$ conservation, and hence independent switch was not required. In version (v2.07f), the compatibility with these previous versions is ensured in the following way. If switch ISIMTY is set to $-1$ (the default value), than the code resets its value to ITIREV, see Sec. 3.2, and stops in case switch ISIMPY is not set to 1 (the default value). In this mode, the present version works in exactly the same way as the former versions. On the other hand, if switch ISIMTY is set to 0 or 1, then the values of ISIMPY and ITIREV must correspond to one of the five options allowed by the group structure, and enumerated in Table 2, otherwise the code stops. Note that in the present version we introduced switch ITIREV=1 or 0 (time-reversal conserved or not) as a convenient replacement of the value 1$-$IROTAT, where switch IROTAT=0 or 1 (no rotation or rotation) was introduced in Sec. II-3.3.

Table 3: Ternary set of conserved and nonconserved symmetries allowed in the code HFODD version (v2.07f).

| Option | Symmetries | | | Switches | | |
|---|---|---|---|---|---|---|
| | $\hat{R}_y$ | $\hat{S}_x^T$ | $\hat{S}_z^T$ | `ISIGNY` | `ISIMTX` | `ISIMTZ` |
| T1 | conserved | conserved | conserved | 1 | 1 | 1 |
| T2 | conserved | nonconserved | nonconserved | 1 | 0 | 0 |
| T3 | nonconserved | conserved | nonconserved | 0 | 1 | 0 |
| T4 | nonconserved | nonconserved | conserved | 0 | 0 | 1 |
| T5 | nonconserved | nonconserved | nonconserved | 0 | 0 | 0 |

The third set of symmetries, composed of the $y$-signature ($\hat{R}_y$), $x$-simplex$^T$ ($\hat{S}_x^T$), and $z$-simplex$^T$ ($\hat{S}_z^T$), and switches `ISIGNY`, `ISIMTX`, and `ISIMTZ`, is represented in Table 3. In version (v2.07f) we have unblocked option T2, which was not available in previous versions, see discussion in Sec. I-3.4.

Triples of symmetries listed in Tables 1–3 are linked in two points: the $y$-simplex ($\hat{S}_y$) appears in Tables 1 and 2, and the $y$-signature ($\hat{R}_y$) appears in Tables 1 and 3. Therefore, options P1 and P2 (conserved $y$-simplex) must be linked with options S1 and S2, and the same holds for P3, P4, P5 (nonconserved $y$-simplex) and S3, S4, S5. Similarly, options P1 and P3 (conserved $y$-signature) must be linked with options T1 and T2, and options P2, P4, P5 with T3, T4, T5. Altogether, we obtain 34 allowed options that are illustrated by a tree of links in Fig. 1. They are also listed in Table 4, together with values of the corresponding program switches.

The last column of Table 4 gives generators of conserved point groups associated with each option. The generators are selected by using conventions formulated in Ref. [6] (cf. Table I in this reference). The first option (P1-S1-T1) corresponds to the whole $\mathrm{D}_{2\mathrm{h}}^{\mathrm{T}}$ group conserved, while in the last option (P5-S5-T5), the group contains only the identity operator $\hat{E}$. Altogether, symmetries allowed in the code HFODD cover 17 out of 26 nontrivial subgroups of $\mathrm{D}_{2\mathrm{h}}^{\mathrm{T}}$, which have been enumerated in Ref. [6]. We note here in passing that simple extensions of the available symmetries can be done in the following way:

1. By adding symmetry $\hat{R}_x$ to the present set, one could cover 5 more nontrivial subgroups of $\mathrm{D}_{2\mathrm{h}}^{\mathrm{T}}$, i.e., $\{\hat{R}_x\hat{R}_y\hat{T}\}$, $\{\hat{R}_x\hat{R}_y\hat{P}\}$, $\{\hat{R}_x\hat{R}_y^T\hat{P}^T\}$, $\{\hat{R}_x\hat{S}_y\}$, and $\{\hat{R}_x\hat{R}_y\}$.

2. By adding symmetry $\hat{R}_x^T$ to the present set, one could cover 4 more nontrivial subgroups of $\mathrm{D}_{2\mathrm{h}}^{\mathrm{T}}$, i.e., $\{\hat{R}_x\hat{R}_y\hat{T}\}$, $\{\hat{R}_y\hat{R}_x^T\hat{P}^T\}$, $\{\hat{R}_y\hat{R}_x^T\}$, and $\{\hat{R}_x^T\}$.

3. By adding symmetry $\hat{S}_x$ to the present set, one could cover 3 more nontrivial subgroups of $\mathrm{D}_{2\mathrm{h}}^{\mathrm{T}}$, i.e., $\{\hat{R}_z\hat{R}_x^T\hat{P}^T\}$, $\{\hat{R}_x\hat{R}_y\hat{P}\}$, and $\{\hat{R}_z\hat{S}_x\}$.

Obviously, by adding both $\hat{R}_x$ and $\hat{R}_x^T$, one could cover 7 more nontrivial subgroups of $\mathrm{D}_{2\mathrm{h}}^{\mathrm{T}}$ as compared to the present status of the code, thus altogether covering 24 out of 26 nontrivial subgroups. The missing two would then be only $\{\hat{R}_x^T \hat{P}^T\}$ and $\{\hat{P}^T\}$. The above possible extensions are left for future development, and can easily be implemented provided there will be a physics motivation to study such new cases.

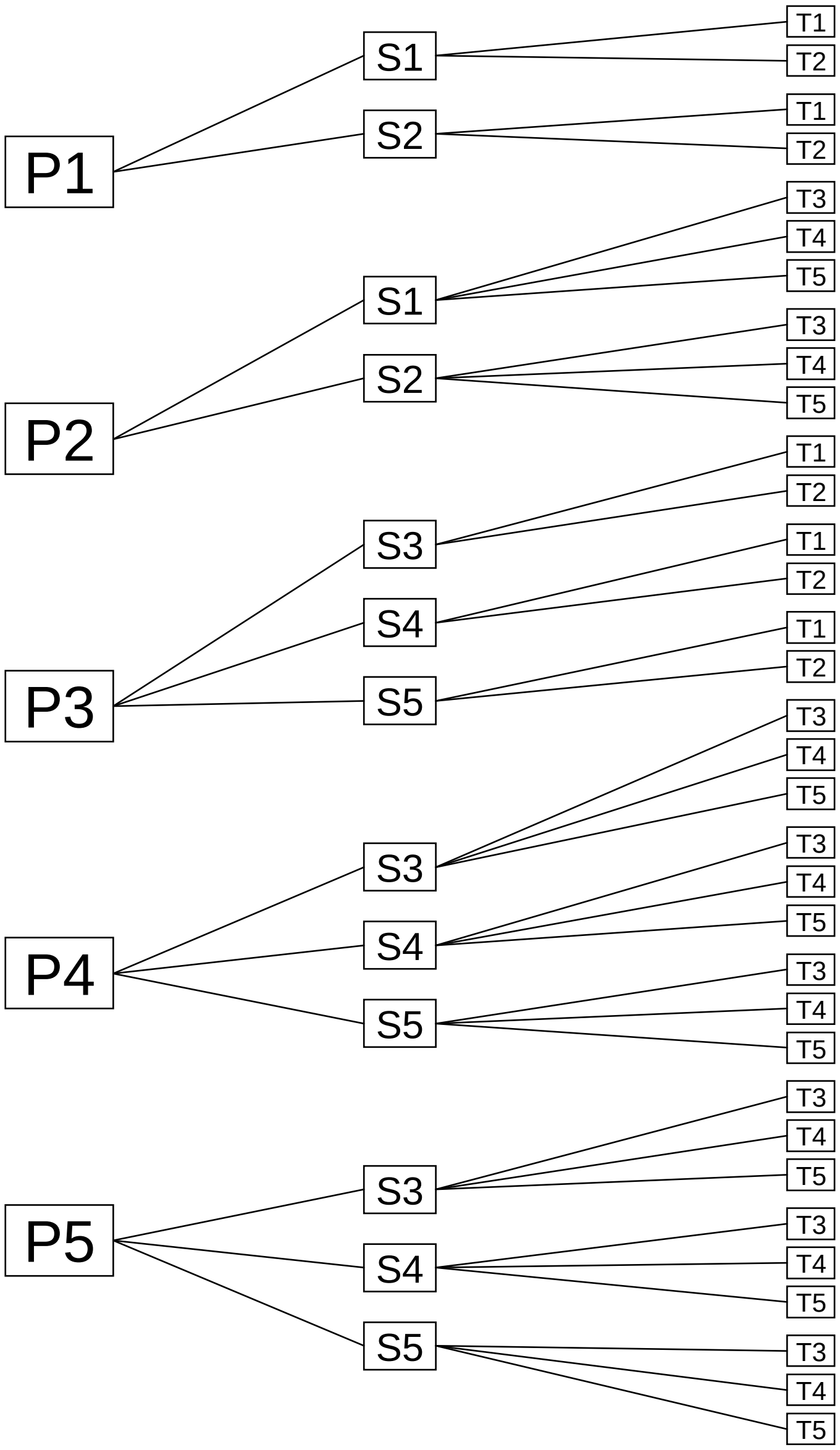

Figure 1: Tree diagram of all allowed options in the code HFODD version (v2.07f). Symbols P, S, and T refer to options listed in Tables 1–3. Lines connect sets of allowed options that are listed in Table 4.

Table 4: Complete list of all conserved (C) or nonconserved (N) symmetries allowed in the code HFODD version (v2.07f). The last column gives generators of the conserved subgroup of $D_{2h}^{T}$.

| Option | Symmetries | | | | | | | Switches | | | | | | | Conserved Group |
|---|---|---|---|---|---|---|---|---|---|---|---|---|---|---|---|
| | $\hat{S}_y$ | $\hat{R}_y$ | $\hat{P}$ | $\hat{T}$ | $\hat{S}_y^T$ | $\hat{S}_x^T$ | $\hat{S}_z^T$ | ISIMPY | ISIGNY | IPARTY | ITIREV | ISIMTY | ISIMTX | ISIMTZ | |
| P1-S1-T1 | C | C | C | C | C | C | C | 1 | 1 | 1 | 1 | 1 | 1 | 1 | $\{\hat{R}_x\hat{R}_y\hat{T}\hat{P}\}$ |
| P1-S1-T2 | C | C | C | C | C | N | N | 1 | 1 | 1 | 1 | 1 | 0 | 0 | $\{\hat{R}_y\hat{T}\hat{P}\}$ |
| P1-S2-T1 | C | C | C | N | N | C | C | 1 | 1 | 1 | 0 | 0 | 1 | 1 | $\{\hat{R}_y\hat{S}_z^T\hat{P}\}$ |
| P1-S2-T2 | C | C | C | N | N | N | N | 1 | 1 | 1 | 0 | 0 | 0 | 0 | $\{\hat{R}_y\hat{P}\}$ |
| P2-S1-T3 | C | N | N | C | C | C | N | 1 | 0 | 0 | 1 | 1 | 1 | 0 | $\{\hat{R}_z\hat{S}_x\hat{T}\}$ |
| P2-S1-T4 | C | N | N | C | C | N | C | 1 | 0 | 0 | 1 | 1 | 0 | 1 | $\{\hat{R}_x\hat{S}_y\hat{T}\}$ |
| P2-S1-T5 | C | N | N | C | C | N | N | 1 | 0 | 0 | 1 | 1 | 0 | 0 | $\{\hat{S}_y\hat{T}\}$ |
| P2-S2-T3 | C | N | N | N | N | C | N | 1 | 0 | 0 | 0 | 0 | 1 | 0 | $\{\hat{S}_y\hat{S}_x^T\}$ |
| P2-S2-T4 | C | N | N | N | N | N | C | 1 | 0 | 0 | 0 | 0 | 0 | 1 | $\{\hat{S}_y\hat{S}_z^T\}$ |
| P2-S2-T5 | C | N | N | N | N | N | N | 1 | 0 | 0 | 0 | 0 | 0 | 0 | $\{\hat{S}_y\}$ |
| P3-S3-T1 | N | C | N | C | N | C | C | 0 | 1 | 0 | 1 | 0 | 1 | 1 | $\{\hat{R}_y\hat{S}_z\hat{T}\}$ |
| P3-S3-T2 | N | C | N | C | N | N | N | 0 | 1 | 0 | 1 | 0 | 0 | 0 | $\{\hat{R}_y\hat{T}\}$ |
| P3-S4-T1 | N | C | N | N | C | C | C | 0 | 1 | 0 | 0 | 1 | 1 | 1 | $\{\hat{R}_x\hat{R}_y\hat{P}^T\}$ |
| P3-S4-T2 | N | C | N | N | C | N | N | 0 | 1 | 0 | 0 | 1 | 0 | 0 | $\{\hat{R}_y\hat{P}^T\}$ |
| P3-S5-T1 | N | C | N | N | N | C | C | 0 | 1 | 0 | 0 | 0 | 1 | 1 | $\{\hat{R}_y\hat{S}_z^T\}$ |
| P3-S5-T2 | N | C | N | N | N | N | N | 0 | 1 | 0 | 0 | 0 | 0 | 0 | $\{\hat{R}_y\}$ |
| P4-S3-T3 | N | N | C | C | N | C | N | 0 | 0 | 1 | 1 | 0 | 1 | 0 | $\{\hat{R}_x\hat{T}\hat{P}\}$ |
| P4-S3-T4 | N | N | C | C | N | N | C | 0 | 0 | 1 | 1 | 0 | 0 | 1 | $\{\hat{R}_z\hat{T}\hat{P}\}$ |
| P4-S3-T5 | N | N | C | C | N | N | N | 0 | 0 | 1 | 1 | 0 | 0 | 0 | $\{\hat{T}\hat{P}\}$ |
| P4-S4-T3 | N | N | C | N | C | C | N | 0 | 0 | 1 | 0 | 1 | 1 | 0 | $\{\hat{R}_z\hat{S}_x^T\hat{P}\}$ |
| P4-S4-T4 | N | N | C | N | C | N | C | 0 | 0 | 1 | 0 | 1 | 0 | 1 | $\{\hat{R}_x\hat{S}_y^T\hat{P}\}$ |
| P4-S4-T5 | N | N | C | N | C | N | N | 0 | 0 | 1 | 0 | 1 | 0 | 0 | $\{\hat{S}_y^T\hat{P}\}$ |
| P4-S5-T3 | N | N | C | N | N | C | N | 0 | 0 | 1 | 0 | 0 | 1 | 0 | $\{\hat{S}_x^T\hat{P}\}$ |
| P4-S5-T4 | N | N | C | N | N | N | C | 0 | 0 | 1 | 0 | 0 | 0 | 1 | $\{\hat{S}_z^T\hat{P}\}$ |
| P4-S5-T5 | N | N | C | N | N | N | N | 0 | 0 | 1 | 0 | 0 | 0 | 0 | $\{\hat{P}\}$ |
| P5-S3-T3 | N | N | N | C | N | C | N | 0 | 0 | 0 | 1 | 0 | 1 | 0 | $\{\hat{S}_x\hat{T}\}$ |
| P5-S3-T4 | N | N | N | C | N | N | C | 0 | 0 | 0 | 1 | 0 | 0 | 1 | $\{\hat{S}_z\hat{T}\}$ |
| P5-S3-T5 | N | N | N | C | N | N | N | 0 | 0 | 0 | 1 | 0 | 0 | 0 | $\{\hat{T}\}$ |
| P5-S4-T3 | N | N | N | N | C | C | N | 0 | 0 | 0 | 0 | 1 | 1 | 0 | $\{\hat{R}_z\hat{S}_x^T\}$ |
| P5-S4-T4 | N | N | N | N | C | N | C | 0 | 0 | 0 | 0 | 1 | 0 | 1 | $\{\hat{R}_x\hat{S}_y^T\}$ |
| P5-S4-T5 | N | N | N | N | C | N | N | 0 | 0 | 0 | 0 | 1 | 0 | 0 | $\{\hat{S}_y^T\}$ |
| P5-S5-T3 | N | N | N | N | N | C | N | 0 | 0 | 0 | 0 | 0 | 1 | 0 | $\{\hat{S}_x^T\}$ |
| P5-S5-T4 | N | N | N | N | N | N | C | 0 | 0 | 0 | 0 | 0 | 0 | 1 | $\{\hat{S}_z^T\}$ |
| P5-S5-T5 | N | N | N | N | N | N | N | 0 | 0 | 0 | 0 | 0 | 0 | 0 | $\{\hat{E}\}$ |

### 2.3 Tilted-axis cranking

The tilted-axis cranking is realized by using in the energy functional the extended cranking term, cf. Eq. (I-23), which now reads

$$\mathcal{E}^{\rm cran} = \sum_{a=x,y,z} \left[ -\omega_{J0a}\langle \hat{J}_{0a}\rangle + C_{Ja}\left(\langle \hat{J}_{0a}\rangle - \bar{J}_{0a}\right)^2 - \omega_{J1a}\langle \hat{J}_{1a}\rangle \right] + C_A \left( \frac{\boldsymbol{\omega}_{J0} \times \langle \hat{\boldsymbol{J}}_0\rangle}{\omega_{J0}} \right)^2 . \qquad (1)$$

The first two terms under the sum are simple generalizations of the standard linear and quadratic spin constraints to three dimensions. Different stiffness constants $C_{Ja}$ are allowed in three Cartesian directions $a{=}x, y, z$. These standard constraints act on the standard isoscalar (total) average angular momentum vector, $\langle \hat{J}_{0a}\rangle{=}\langle \hat{J}_{na}\rangle{+}\langle \hat{J}_{pa}\rangle$, which is a sum of the neutron ($n$) and proton ($p$) component. The third term under the sum constitutes a linear constraint on the isovector average angular momentum vector, $\langle \hat{J}_{1a}\rangle{=}\langle \hat{J}_{na}\rangle{-}\langle \hat{J}_{pa}\rangle$. It is introduced to facilitate the fixing of separate neutron and proton angular momentum vectors, which can be essential when trying to localize, e.g., the shears configurations. Of course, after a proper configuration is found, the final constraints should involve only the isoscalar component of the angular momentum vector. The last term in Eq. (1) constraints to zero the angle between the angular frequency and angular momentum vectors. Since in the self-consistent solutions this angle must be equal to zero [7], the last constraint helps to reach the self-consistent solution faster. However, since it is build upon the angular frequency vector pertaining to the linear isoscalar constraint, it cannot be used neither in conjunction with the quadratic nor with the isovector constraint.

In practical calculations, it turns out that the angle between the angular frequency and angular momentum converges to zero extremely slowly. This is so because for an angular frequency vector fixed in space, the whole nucleus must turn in space in order to align its angular momentum with the angular frequency. Therefore, a much faster procedure is to proceed in an opposite way, i.e., in each iteration to force the angular frequency to be aligned or anti-aligned with the current angular momentum vector (see switch `IMOVAX` in Sec. 3.5). This is a purely heuristic procedure, because it does not correspond to a minimization of any given Routhian. However, once the self-consistent solution is found (the angular frequency and angular momentum vectors become parallel or antiparallel) it is the Routhian for the final angular frequency which has taken the minimum value.

### 2.4 Multipole, surface multipole, and magnetic moments

Previous versions of the code worked in the regime of the conserved $y$-simplex, which ensured that all components of all multipole moments were real, cf. Eq. (I-56). Since in the present version the $y$-simplex symmetry was released, we have to carefully define all phase conventions used in calculating matrix elements of multipole operators. This is done by adopting the following convention for multipole moments:

$$Q_{\lambda\mu}(\boldsymbol{r}) = a_{\lambda\mu} r^{\lambda} Y^{*}_{\lambda\mu}(\theta,\phi), \qquad (2)$$

where $Y_{\lambda\mu}$ are the standard spherical harmonics in the convention of Ref. [8]. Note that the complex-conjugate spherical harmonics enter the adopted definition of multipole moments, i.e., the phase convention corresponds to that used in electrodynamics, see Eq. (4.3) of Ref. [9].

Real factors $a_{\lambda\mu}$=$a_{\lambda,-\mu}$ ensure a traditional normalization of the low-multipolarity moments (for $\lambda\leq2$), such that they correspond to moments of simple polynomials of coordinates, see Table 5. For $\lambda>2$ we set $a_{\lambda\mu}$=1. We have kept the factor of $\sqrt{3}$ in the definition of $Q_{22}$ in order to conform to the standard definition of the Bohr $\gamma$ deformation [10], i.e., $\tan\gamma=\Re\langle Q_{22}\rangle/\langle Q_{20}\rangle$. As usual, moments for negative values of the magnetic quantum number $\mu$ can be obtained from $Q_{\lambda,-\mu}=(-1)^{\mu}Q^{*}_{\lambda\mu}$. Whenever the multipole moments become complex, the code prints the real parts of non-negative-$\mu$ components and imaginary parts of negative-$\mu$ components.

Table 5: Adopted definitions of the normalization factors $a_{\lambda\mu}$ and the corresponding multipole moments $Q_{\lambda\mu}$.

| $\lambda$ | $\mu$ | $a_{\lambda\mu}$ | $Q_{\lambda\mu}$ |
|---|---|---|---|
| 0 | 0 | $\sqrt{4\pi}$ | 1 |
| 1 | 0 | $\sqrt{4\pi/3}$ | $z$ |
| 1 | 1 | $-\sqrt{8\pi/3}$ | $x-iy$ |
| 2 | 0 | $\sqrt{16\pi/5}$ | $2z^2-x^2-y^2$ |
| 2 | 1 | $-\sqrt{8\pi/15}$ | $zx-izy$ |
| 2 | 2 | $\sqrt{32\pi/5}$ | $\sqrt{3}(x^2-y^2-2ixy)$ |

Since the adopted normalization factors depend on the magnetic quantum number $\mu$, rotational invariants *are not equal* to sums of moduli squared of all magnetic components for a given multipole moment. In fact, it is easy to check that for $\lambda$=1 and 2 the invariant combinations are $|Q_{10}|^2+|Q_{11}|^2$ and $|Q_{20}|^2+12|Q_{21}|^2+|Q_{22}|^2$.

Apart from calculating multipole moments, the HFODD code version (v2.07f) also calculates the average values of the so-called surface multipole moments,

$$Q^S_{\lambda\mu}(\boldsymbol{r}) = a_{\lambda\mu}r^{\lambda+2}Y^{*}_{\lambda\mu}(\theta,\phi), \tag{3}$$

which are much more sensitive to matter distribution in the surface region than are the standard multipole moments (2). Moreover, the surface dipole moment is needed to calculate the Schiff moment of a nucleus, see, e.g., Ref. [11]. Note that within the chosen normalization, the surface monopole moment is equal to the radius squared, $Q^S_{00}=r^2$. Values of the surface multipole moments can be constrained in the same way as the values of the multipole moments, so the multipole constraint term (I-22) now takes the form

$$\mathcal{E}^{\text{mult}} = \sum_{\lambda\mu} C_{\lambda\mu}\left(\langle\hat{Q}_{\lambda\mu}\rangle-\bar{Q}_{\lambda\mu}\right)^2 + \sum_{\lambda\mu} C^S_{\lambda\mu}\left(\langle\hat{Q}^S_{\lambda\mu}\rangle-\bar{Q}^S_{\lambda\mu}\right)^2. \tag{4}$$

The new version also calculates the average values of the magnetic moment operators [10],

$$\hat{M}_{\lambda\mu} = \left(g_s\hat{\boldsymbol{S}} + \tfrac{2}{\lambda+1}g_l\hat{\boldsymbol{L}}\right)\cdot\boldsymbol{\nabla}\left[r^{\lambda}Y^{*}_{\lambda\mu}(\theta,\phi)\right], \tag{5}$$

where $\hat{\boldsymbol{S}}$ and $\hat{\boldsymbol{L}}$ are the spin and orbital angular momentum operators, respectively, and $g_s$ and $g_l$ are the standard gyromagnetic factors. Constraints on magnetic moments are not yet implemented.

Multipole, surface multipole, and magnetic moments are calculated only for those multipolarities $\lambda$ and magnetic components $\mu$ which are allowed for a given pattern of conserved symmetries, see Ref. [5]. Whenever a shift of the center of mass from the origin of the coordinate frame is allowed, the code also calculates multipole and surface multipole moments in the center-of-mass reference frame. Calculation of the magnetic moments in the center-of-mass reference frame is not yet implemented.

Considering a rotation of the principal axes of the mass distribution with respect to the center-of-mass reference frame, the code determines the Euler angles corresponding to such a rotation, and then calculates all moments and angular momenta in the principal-axes (intrinsic) reference frame. The Euler angles are determined by diagonalizing the Cartesian matrix of the quadrupole tensor, see Ref. [8], and thus finding the orthogonal 3×3 transformation that corresponds to the required rotation. The order of eigenvalues is chosen in such way that in the principal-axes reference frame one has $\langle Q_{yy}\rangle\leq\langle Q_{xx}\rangle\leq\langle Q_{zz}\rangle$, where $Q_{aa}$=$3x_a^2-r^2$, or equivalently $\langle y^2\rangle\leq\langle x^2\rangle\leq\langle z^2\rangle$. Therefore, the obtained Euler angles bring the system to the first sector $0^\circ \leq \gamma \leq 60^\circ$ of the standard Bohr $\gamma$ deformation [10].

Note that the principal-axes (intrinsic) reference frame defined in such a way can differ from the original frame even in the case of conserved signature and parity, i.e, when the system has three symmetry planes. Indeed, in such a case, the Euler angles may correspond to a rotation that simply exchanges names and directions of the Cartesian axes. Since this need not be an interesting transformation, the user can manually switch off the printing of results that pertain to the intrinsic frame, by using switch `INTRIP`, see Sec. 3.6.

Note that the center-of-mass and principle-axes reference frames are determined from the mass dipole and quadrupole moments, respectively, and hence the neutron and proton distributions need not be individually brought to their respective center-of-mass and principal-axes frames. In other words, in the center-of-mass reference frame the neutron and proton dipole moments can be different from zero, and in the principle-axes reference frame the neutron and proton quadrupole moments need not obey conditions $\langle Q_{21}\rangle$=0 and $\Im\langle Q_{22}\rangle$=0. Note also that constraints on the multipole and surface multipole moments are formulated with respect to moments calculated in the original reference frame.

### 2.5 Solution of the Hartree-Fock-Bogolyubov equations

In order to incorporate pairing correlations for rotating states, the new version (v2.07f) of the code HFODD solves the standard HFB equation [10],

$$\begin{pmatrix} h' - \lambda & \Delta \\ -\Delta^* & -h'^* + \lambda \end{pmatrix} \begin{pmatrix} A \\ B \end{pmatrix} = \begin{pmatrix} A \\ B \end{pmatrix} E, \tag{6}$$

by using the Cartesian HO basis. Here, $h'$ is the s.p. Routhian operator (I-26), with the cranking term $-\boldsymbol{\omega}\cdot\hat{\boldsymbol{J}}_0$ generalized to three dimensions, $\lambda$ is the Fermi energy (the HFB equation is solved separately for protons and neutrons), $\Delta$ is the antisymmetric pairing potential, $E$ is a diagonal matrix of quasiparticle energies, and $A$ and $B$ are the upper and lower components of the quasiparticle wavefunctions.

For the conserved simplex symmetry, which in the present version is assumed when solving the HFB equation, the Routhian and the pairing potential acquire the following block forms [12, 10]:

$$h' = \begin{pmatrix} h'_+ & 0 \\ 0 & h'_- \end{pmatrix} \quad , \quad \Delta = \begin{pmatrix} 0 & \Delta_+ \\ \Delta_- & 0 \end{pmatrix}, \tag{7}$$

where the first and second halfs of the basis states correspond to positive and negative simplex, respectively. Therefore, the HFB equation decouples into two independent equations, of which only one needs to be solved. The code HFODD solves the equation

$$\begin{pmatrix} h'_+ - \lambda & \Delta_+ \\ -\Delta^*_- & -h'^*_- + \lambda \end{pmatrix} \begin{pmatrix} A_+ & B^*_+ \\ B_- & A^*_- \end{pmatrix} = \begin{pmatrix} A_+ & B^*_+ \\ B_- & A^*_- \end{pmatrix} \begin{pmatrix} E_- & 0 \\ 0 & -E_+ \end{pmatrix}, \tag{8}$$

from which the complete solution of Eq. (6) is reconstructed as

$$A = \begin{pmatrix} 0 & A_+ \\ A_- & 0 \end{pmatrix} \quad , \quad B = \begin{pmatrix} B_+ & 0 \\ 0 & B_- \end{pmatrix} \quad , \quad E = \begin{pmatrix} E_+ & 0 \\ 0 & E_- \end{pmatrix}. \tag{9}$$

Note that if the s.p. space contains $M$ states ($M/2$ in each of the two simplexes $s{=}\pm i$), then the HFB equation (6) has a dimension of $2M\times 2M$, matrices without the $\pm$ indices ($h'$, $\Delta$, $A$, $B$, ...) have a dimension of $M\times M$, and matrices with the $\pm$ indices ($h'_\pm$, $\Delta_\pm$, $A_\pm$, $B_\pm$, ...) have a dimension of $M/2\times M/2$. Hence, equation (8), which the code solves, has the dimension $M\times M$ which is twice smaller than the complete HFB equation (6). Note also that we use the convention of indices $\pm$ such that they correspond to simplexes of the *left* matrix indices.

For the conserved time-reversal symmetry, diagonal matrices of eigenvalues $E_+$ and $E_-$ are identical to one another, and the separation of eigenvectors of Eq. (8) into two simplexes is trivial; it is enough to group positive and negative quasiparticle energies together. For broken time-reversal this procedure does not work, because there can be more than half positive or negative quasiparticle energies in the spectrum of Eq. (8). So in general one has to put the half of largest eigenvalues into matrix $E_-$, and the half of smallest into matrix $E_+$, irrespective of their signs. Such a choice leads to the solution that corresponds to the so-called quasiparticle vacuum.

Based on the solution of the matrix equation (8), we have upper and lower components of the quasiparticle wave functions in the space coordinates as

$$\varphi^{(1)}_{\alpha,s=\pm i}(\boldsymbol{r}\sigma) = 2\sigma \sum_{\boldsymbol{n}} \psi^*_{\boldsymbol{n},s=\pm i}(\boldsymbol{r},-\sigma) A^*_{\pm,\boldsymbol{n}\alpha}, \tag{10a}$$

$$\varphi^{(2)}_{\alpha,s=\pm i}(\boldsymbol{r}\sigma) = \sum_{\boldsymbol{n}} \psi_{\boldsymbol{n},s=\pm i}(\boldsymbol{r}\sigma) B^*_{\pm,\boldsymbol{n}\alpha}, \tag{10b}$$

where $\psi_{\boldsymbol{n},s=\pm i}(\boldsymbol{r}\sigma)$ are the HO simplex wave functions (I-78) in space coordinates (I-76) and $\boldsymbol{n}{=}(n_x, n_y, n_z)$ are the numbers of the HO quanta in three Cartesian directions. In Eq. (10) we have used the fact that the s.p. basis states of either of the two simplexes, $s = \pm i$, can be numbered by the HO quantum numbers $\boldsymbol{n}$. We have also introduced index $\alpha{=}1,\ldots,M/2$, which numbers eigenstates of Eq. (8) in both "halfs" of the spectrum defined above.

From the quasiparticle wave functions we obtain the standard particle and pairing density matrices [13],

$$\rho(\boldsymbol{r}\sigma, \boldsymbol{r}'\sigma') = \sum_\alpha \left( \varphi^{(2)}_{\alpha,s=+i}(\boldsymbol{r}\sigma)\varphi^{(2)*}_{\alpha,s=+i}(\boldsymbol{r}'\sigma') + \varphi^{(2)}_{\alpha,s=-i}(\boldsymbol{r}',\sigma')\varphi^{(2)*}_{\alpha,s=-i}(\boldsymbol{r},\sigma') \right), \tag{11a}$$

$$\tilde\rho(\boldsymbol{r}\sigma,\boldsymbol{r}'\sigma') = -\sum_\alpha \left(\varphi^{(2)}_{\alpha,s=+i}(\boldsymbol{r}\sigma)\varphi^{(1)*}_{\alpha,s=-i}(\boldsymbol{r}'\sigma') + 4\sigma\sigma'\varphi^{(2)}_{\alpha,s=+i}(\boldsymbol{r}',-\sigma')\varphi^{(1)*}_{\alpha,s=-i}(\boldsymbol{r},-\sigma')\right), \quad (11b)$$

where the sum over $\alpha$ is performed up to the maximum equivalent-spectrum energy $\bar{e}_{\max}$, see Ref. [14] for details. All particle-hole mean-field potentials can be calculated from the particle density matrix $\rho(\boldsymbol{r}\sigma,\boldsymbol{r}'\sigma')$ and its derivatives (see I), while the particle-particle mean-field potentials can be calculated from the pairing density matrix $\tilde\rho(\boldsymbol{r}\sigma,\boldsymbol{r}'\sigma')$ [13]. In the present implementation of the code HFODD, terms depending on derivatives of the particle-particle density matrix are not taken into account, and hence the pairing potential depends only on the local pair density

$$\tilde\rho(\boldsymbol{r}) = -2\sum_{\alpha\sigma}\varphi^{(2)}_{\alpha,s=+i}(\boldsymbol{r}\sigma)\varphi^{(1)*}_{\alpha,s=-i}(\boldsymbol{r}\sigma), \quad (12)$$

i.e.,

$$\tilde h(\boldsymbol{r}) = \tfrac{1}{2}f(\boldsymbol{r})\tilde\rho(\boldsymbol{r}), \quad (13)$$

where $f(\boldsymbol{r})$ is the standard formfactor of the zero-range density-dependent pairing force,

$$f(\boldsymbol{r}) = V_0 + V_1\rho^\alpha(\boldsymbol{r}) = V_0\left(1-\left(\frac{\rho(\boldsymbol{r})}{\rho_0}\right)^\alpha\right). \quad (14)$$

Note that an attractive interaction requires $V_0<0$.

Finally, matrix elements of the pairing potential in the simplex HO basis, which are needed in Eq. (8), can be calculated in exactly the same way as matrix elements of the central mean-field potential, Sec. I-4.2, i.e.,

$$\Delta_{\pm,\boldsymbol{n}\boldsymbol{n}'} = \pm F(n_y-n'_y)\int \mathrm{d}^3\boldsymbol{r}\,\psi^*_{\boldsymbol{n}}(\boldsymbol{r})\tilde h(\boldsymbol{r})\psi_{\boldsymbol{n}'}(\boldsymbol{r}) \quad (15)$$

where factors $F(n)$ are defined as in Eq. (I-85),

$$F(n) = \begin{cases} (-1)^{\frac{n}{2}} & \text{for } n \text{ even},\\ 0 & \text{for } n \text{ odd}.\end{cases}, \quad (16)$$

and $\psi_{\boldsymbol{n}}$ are the HO basis states.

## 2.6 Spin-orbit energy density

A complete isospin dependence of the spin-orbit term [15] has already been implemented in several standard Skyrme forces. Within such a generalized parametrization, the isoscalar and isovector coupling constants, which define the spin-orbit term, see Eq. (I-12), read

$$C_0^{\nabla J} = C_0^{\nabla j} = -\tfrac{1}{2}W_0 - \tfrac{1}{4}W'_0, \quad (17a)$$

$$C_1^{\nabla J} = C_1^{\nabla j} = -\tfrac{1}{4}W'_0. \quad (17b)$$

while the traditional parametrization requires that $W'_0 \equiv W_0$. Here, $W_0$ is the strength of the two-body spin-orbit part of the Skyrme force [16, 17]. The corresponding parameters used in Ref. [15] are $b_4{=}\tfrac{1}{2}W_0$ and $b'_4{=}\tfrac{1}{2}W'_0$.

The code HFODD uses the energy density coupling constants, and not the parameters of the Skyrme force, so generalization (17) has always been available through appropriate scaling of the coupling constants, see Sec. II-3.2. In the new version (v2.07f), the use of the above generalization was facilitated by introducing explicit input data parameters that allow to directly handle the strengths $W'_0$ and $W_0$.

### 2.7 Center-of-mass correction to the energy

Since the HF or HFB states break the translational symmetry, one has to add, in principle, to the total energy the so-called center-of-mass correction [18],

$$E_{\text{c.m.}} = -\frac{1}{2mA}\langle \hat{\boldsymbol{P}}^2 \rangle, \tag{18}$$

where $\hat{\boldsymbol{P}}{=}-i\hbar\sum_{i=1}^{A}\nabla_i$ is the total linear momentum operator. Since evaluation of this correction is time consuming, one often uses the approximation which keeps only the direct term, i.e.,

$$E_{\text{c.m.}} \simeq E_{\text{c.m.}}^{\text{dir}} = -\frac{1}{A}\langle \hat{T} \rangle, \tag{19}$$

where $\hat{T}{=}-\frac{\hbar^2}{2m}\sum_{i=1}^{A}\nabla_i^2$ is the one-body kinetic-energy operator. Within this approximation, a simple renormalization of the nucleon mass,

$$\frac{1}{m} \longrightarrow \frac{1}{m}\left(1-\frac{1}{A}\right), \tag{20}$$

allows to include correction (19) *before variation.* This way of proceeding is traditionally most often employed, and it has also been implemented by default in previous versions of the code HFODD. A more advanced, while still not-too-expensive method consists in evaluating correction (18) *after variation*, i.e., after having performed the HF or HFB iterations. Such an option has now been implemented in version (v2.07f).

### 2.8 Landau parameters

The time-odd coupling constants of the Skyrme energy functional, see I, are poorly known, because of the lack of experimental data that would be sufficiently sensitive to this channel of energy density. Recently, such sensitivity has been analyzed for the effects related to the Gamow-Teller strength in nuclei [19]. In this analysis, it turned out that one can make a useful unique link between some time-odd coupling constants and the Landau parameters $g_0$, $g_0'$, $g_1$, and $g_1'$. In view of that, it is more practical, and physically more intuitive, to use the Landau parameters at saturation density as input data for the code HFODD. This is done by employing Eq. (15) of Ref. [19], together with ratios of density-dependent coupling constants, $C_0^s[\rho]$ and $C_1^s[\rho]$, i.e.,

$$\begin{aligned}
x_0 &= C_0^s[0]/C_0^s[\rho_{\text{sat}}], && \text{(21a)}\\
x_1 &= C_1^s[0]/C_1^s[\rho_{\text{sat}}], && \text{(21b)}\\
g_0 &= N_0(2C_0^s[\rho_{\text{sat}}] + 2C_0^T\,\beta\,\rho_{\text{sat}}^{2/3}), && \text{(21c)}\\
g_0' &= N_0(2C_1^s[\rho_{\text{sat}}] + 2C_1^T\,\beta\,\rho_{\text{sat}}^{2/3}), && \text{(21d)}\\
g_1 &= -2N_0\,C_0^T\,\beta\,\rho_{\text{sat}}^{2/3}, && \text{(21e)}\\
g_1' &= -2N_0\,C_1^T\,\beta\,\rho_{\text{sat}}^{2/3}, && \text{(21f)}
\end{aligned}$$

where $\beta = (3\pi^2/2)^{2/3}$ and

$$N_0 = \pi^{-2}\left(\frac{\hbar^2}{2m}\right)^{-1}\left(\frac{m^*}{m}\right)\left(\frac{3\pi^2\rho_{\rm sat}}{2}\right)^{1/3}. \qquad (22)$$

Altogether, six parameters appearing at the left-hand-sides of Eqs. (21) uniquely determine six coupling constants $C_0^s[0]$, $C_0^s[\rho_{\rm sat}]$, $C_1^s[0]$, $C_1^s[\rho_{\rm sat}]$, $C_0^T$, and $C_1^T$, provided the constant $\hbar^2/2m$, effective mass $m^*/m$, and saturation density $\rho_{\rm sat}$ are given.

### 2.9 Diagonalization subroutines

As described in Section 5.1, an interface to the LAPACK subroutine ZHPEVX has been created. This allows using public-domain diagonalization subroutines, as an alternative of using the NAGLIB or ESSL packages, previously described in Sect. II-5.3, and allows faster calculations as compared to the previously used subroutine ZHPEV, Sect. III-5.1.

### 2.10 Memory management

As described in Section 5.2, memory management that is based on memory-allocation features available within FORTRAN-90, has been implemented. In the standard form, the source code is using only the FORTRAN-77 commands, however, if the memory size becomes an issue and the FORTRAN-90 compiler is available, the source code can be easily transformed into the FORTRAN-90 form.

## 3 Input data file

Structure of the input data file has been described in II; in version (v2.07f) of the code HFODD this structure is exactly the same. All previous items of the input data file remain valid, and several new items were added, as described in Secs. 3.1–3.8.

Together with the FORTRAN source code in the file `hfodd.f`, several examples of the input data files are provided. File `dy152-e.dat` contains all the valid input items, and the input data are identical to the default values. Therefore, the results of running the code with the input data file `dy152-e.dat` are identical to those obtained for the input data file containing only one line with the keyword `EXECUTE`.

File `la132-a.dat` is reproduced in section TEST RUN INPUT below. It defines two consecutive runs of the code HFODD that lead to a planar solution in $^{132}$La that is oriented in space in such a way that the angular frequency vector of length $\hbar|\boldsymbol{\omega}|$=0.25 MeV has all three Cartesian components not equal to zero. File `la132-a.out` containes results of executing code HFODD version (v2.07f) for the input file `la132-a.dat`. Selected lines from file `la132-a.out` are reproduced in section TEST RUN OUTPUT below.

### 3.1 Interaction

**Keyword:** `SKYRME-STD`

0, 1, 0, 0, 0 = `ISTAND,KETA_J,KETA_W,KETACM,KETA_M`

Parameters of several standard Skyrme forces are encoded within the program. Calculation for a given standard force can be requested by specifying its acronym in the input data file, see Sec. II-3.2. In version (v2.07f), valid acronyms are SIII, SV, SKM*, SKP, SKMP, SKI1, SKO, SKOP, SLY4, SLY5, MSK1, MSK2, MSK3, MSK4, MSK5, MSK6, SKX, SKXC. Along with the force parameters, for each force there is encoded information on how the given force should be used, i.e., with which value of the parameter $\hbar^2/2m$, and with which treatment of tensor, spin-orbit, and center-of-mass terms. For `ISTAND`=1, calculations will be performed with these features set in the way specific for the given force, and the rest of the switches read on the same line will be ignored. For `ISTAND`=0, the switches will define nonstandard features in the following way:

- For `KETA_J`=1 the code will take into account in the functional, and for `KETA_J`=0 will neglect the so-called tensor $J^2$ terms. The second option is equivalent to setting in Eq. (I-12) coupling constants $C_t^J$=0 and $C_t^T$=0.
- For `KETA_W`=0 or 1 the code will use the traditional ($W_0' \equiv W_0$) or generalized ($W_0' \neq W_0$) spin-orbit term, respectively, see Eq. (17). For forces that use generalized spin-orbit term, option `KETA_W`=0 sets $W_0'$ equal to $W_0$, while for forces that use traditional spin-orbit, option `KETA_W`=1 has no effect. For `KETA_W`=2, strengths $W_0$ and $W_0'$ are set equal to the values of `W0_INP` and `W0PINP`, respectively, which are read under the keyword of `SPIN_ORBIT`, see below. Note that if `KETA_W`=2 is used, and keyword `SPIN_ORBIT` is not specified in the input data file, the default values of `W0_INP` and `W0PINP` will supersede those that are encoded within the program for the given Skyrme force.
- For `KETACM`=0 or 1 the code will use the traditional one-body center-of-mass correction (19) before variation or the two-body center-of-mass correction (18) after variation, respectively. Value of `KETACM`=2 is reserved for a future implementation of the two-body center-of-mass correction before variation. For `KETACM`=3 the center-of-mass correction will be neglected.
- For `KETA_M`=0 the code will use the value of $\hbar^2/2m$=20.73620941 MeV fm$^2$, which was encoded in version (v1.75r). For `KETA_M`=1 the code will use the value specific for the given Skyrme force. For `KETA_M`=2 the code will use the value specified in the input data file under keyword `HBAR2OVR2M`, see below.

**Keyword:** `HBAR2OVR2M`

20.73620941 = `HBMINP`

Value of the $\hbar^2/2m$ parameter, which will be used if `KETA_M`=2 is set under keyword `SKYRME-STD`, see above, and ignored otherwise.

**Keyword:** `SPIN_ORBIT`

120.0, 120.0 = `W0_INP,W0PINP`

Strengths $W_0$ and $W_0'$ of the generalized spin-orbit interaction (17), which will be used if `KETA_W`=2 is set under keyword `SKYRME-STD`, see above, and ignored otherwise.

**Keyword:** `LANDAU`

0, 0.0, 0.0, 0.0, 0.0, 0.0, 0.0 =`LANODD,X0_LAN,X1_LAN,`
`G0_LAN,G0PLAN,`
`G1_LAN,G1PLAN`

Three-digit steering switch `LANODD`, followed by the values of $x_0$, $x_1$, $g_0$, $g_0'$, $g_1$, and $g_1'$, see Eq. (21). For `LANODD`=000, definition of coupling constants from the Landau parameters is not used, and the rest of parameters read on the same data line is ignored. For `LANODD`=111, Eq. (21) is solved and coupling constants $C_0^s[0]$, $C_0^s[\rho_{\rm sat}]$, $C_1^s[0]$, $C_1^s[\rho_{\rm sat}]$, $C_0^T$, and $C_1^T$ are determined from the values of $x_0$, $x_1$, $g_0$, $g_0'$, $g_1$, and $g_1'$. For other values of `LANODD`, the following three steps are performed in sequence:

1. *Step one:* If the rightmost digit of `LANODD` is equal to 1 then the coupling constants $C_0^T$ and $C_1^T$ are determined form $g_1$ and $g_1'$, Eqs. (21e) and (21f); otherwise these two coupling constants are determined from the Skyrme force parameters.
2. *Step two:* If the middle digit of `LANODD` is equal to 1 then the coupling constants $C_0^s[\rho_{\rm sat}]$ and $C_1^s[\rho_{\rm sat}]$ are determined form $g_0$ and $g_0'$, Eqs. (21c) and (21d); otherwise these two coupling constants are determined from the Skyrme force parameters.
3. *Step three:* If the leftmost digit of `LANODD` is equal to 1 then the coupling constants $C_0^s[0]$ and $C_1^s[0]$ are determined form $x_0$ and $x_1$, Eqs. (21a) and (21b); otherwise these two coupling constants are determined from the Skyrme force parameters.

All combinations of zeros and ones are allowed in `LANODD`.

**Keyword:** `LANDAU-SAT`

−1.0, −1.0, −1.0 = `HBMSAT,RHOSAT,EFFSAT`

Values of parameters $\hbar^2/2m$, $m^*/m$, and $\rho_{\rm sat}$, respectively, which will be used when solving Eq. (21). If a negative number is read for any of these parameters, then the program will use the corresponding value calculated from parameters of the given Skyrme force.

**Keyword:** `INI_FERMI`

−8.0, −8.0 = `FERINI(0),FERINI(1)`

**Keyword:** `INI_DELTA`

1.0, 1.0 = `DELINI(0),DELINI(1)`

For `IPCONT`=0, see Sec. 3.8, calculations that are restarted from previously saved results will be performed with new values of the neutron and proton Fermi energies, `FERINI(0) and FERINI(1)`, and pairing gaps, `DELINI(0) and DELINI(1)`. New values of pairing gaps are implemented by overwriting the old pairing potentials with constant values of `DELINI(0) and DELINI(1)` for neutrons and protons, respectively. These constant potentials are ignored after the first iteration, i.e., in the first iteration, the mixing of previous and calculated potentials (see Sec. 3.4) is not performed. A possibility of restarting calculations with nonzero pairing is very useful in case the pairing would have vanished in a former run.

**Keyword:** `FIXDELTA_N`

1.0, 0 = `DELFIN,IDEFIN`

For `IDEFIN`=1, pairing calculations will be performed with a fixed value of the neutron pairing gap equal to `DELFIN`. For `IDEFIN`=0, value of `DELFIN` will be ignored.

**Keyword:** `FIXDELTA_P`

1.0, 0 = `DELFIP,IDEFIP`

Same as above but for the proton pairing gap.

**Keyword:** `PAIRNFORCE`

−200.0, 0.0, 1.0 = `PRHO_N,PRHODN,POWERN`

Parameters $V_0$, $V_1$, and $\alpha$, respectively, of the zero-range density-dependent pairing force (14) for neutrons. In case values of $V_1$ and $\alpha$ allow it, the code calculates the value of $\rho_0$ that gives the equivalent form of the formfactor (14). In case $\rho_0$ cannot be calculated, the codes set its value to 1; $\rho_0$ is calculated only for the purpose of information, while internally the code uses only the value of $V_1$.

**Keyword:** `PAIRPFORCE`

−200.0, 0.0, 1.0 = `PRHO_P,PRHODP,POWERP`

Same as above but for the proton pairing force.

**Keyword:** `PAIR_FORCE`

−200.0, 0.0, 1.0 = `PRHO_T,PRHODT,POWERT`

Same as above but for the neutron *and* proton pairing force. This keyword is equivalent to using the above two keywords simultaneously with identical parameters for neutrons and protons.

**Keyword:** `PAIRNINTER`

−200.0, 0.16, 1.0 = `PRHO_N,PRHOSN,POWERN`

Parameters $V_0$, $\rho_0$, and $\alpha$, respectively, of the zero-range density-dependent pairing force (14) for neutrons. In case values of $\rho_0$ and $\alpha$ allow it, the code calculates the value of $V_1$ that gives the equivalent form of the formfactor (14). In case $V_1$ cannot be calculated, the code stops.

**Keyword:** `PAIRPINTER`

−200.0, 0.16, 1.0 = `PRHO_P,PRHOSP,POWERP`

Same as above but for the proton pairing force.

**Keyword:** `PAIR_INTER`

−200.0, 0.16, 1.0 = `PRHO_T,PRHOST,POWERT`

Same as above but for the neutron *and* proton pairing force. This keyword is equivalent to using the above two keywords simultaneously with identical parameters for neutrons and protons.

**Keyword:** `CUTOFF`

60.0 = `ECUTOF`

Cutoff energy $\bar{e}_{\max}$ for summing up contributions of quasiparticle states to density matrices (11), see Sec. 2.5.

### 3.2 Symmetries

**Keyword:** `SIMPLEXY`

1 = `ISIMPY`

Calculation with $y$-simplex conserved (see Sec. 2.2) will be performed for `ISIMPY`=1, while the simplex will be broken for `ISIMPY`=0. Value of `ISIMPY` must be consistent with switches `IPARTY`, `ISIGNY`, `ISIMTY`, and `IROTAT`, see Tables 1 and 2.

**Keyword:** `PARITY`

$-1$ = `IPARTY`

Calculation with parity conserved (see Sec. 2.2) will be performed for `IPARTY`=1, while the parity will be broken for `IPARTY`=0. Value of `IPARTY` must be consistent with switches `ISIMPY` and `ISIGNY`, see Table 1. For `IPARTY`=$-1$ (the compatibility mode), the code sets `IPARTY`=`ISIGNY` and requires that `ISIMPY`=1

**Keyword:** `TIMEREVERS`

0 = `ITIREV`

Calculation with time-reversal conserved (see Sec. 2.2) will be performed for `ITIREV`=1, while this symmetry will be broken for `ITIREV`=0. This switch is used only as a convenient replacement of switch `IROTAT`; `ITIREV`=1 is equivalent to `IROTAT`=0 and `ITIREV`=0 is equivalent to `IROTAT`=1, cf. Sec. II-3.3.

**Keyword:** `TSIMPLEX_Y`

$-1$ = `ISIMTY`

Calculation with $y$-simplex$^T$ conserved (see Sec. 2.2) will be performed for `ISIMTY`=1, while this symmetry will be broken for `ISIMTY`=0. Value of `ISIMTY` must be consistent with switches `ISIMPY` and `IROTAT`, see Table 2. For `ISIMTY`=$-1$ (the compatibility mode), the code sets `ISIMTY` to 1-`IROTAT` and requires that `ISIMPY`=1.

**Keyword:** `TSIMPLEX3D`

1, $-1$, 1 = `ISIMTX,ISIMTY,ISIMTZ`

This keyword allows to simultaneously input all the three switches that define the three $T$-simplexes. It is equivalent to using keywords `TSIMPLEXES` (Sec. II-3.3) and `TSIMPLEX_Y` together. Values of `ISIMTX` and `ISIMTZ` must be consistent with switch `ISIGNY`, see Table 3.

**Keyword:** `HFB`

0 = `IPAHFB`

Calculation with pairing correlations included by solving the HFB equation (see Sec. 2.5) will be performed for `IPAHFB`=1. `IPAHFB`=1 requires `IPAIRI`=1, see Sec. II-3.3. `IPAHFB`=1 also requires `ISIMPY`=1, i.e., for broken simplex symmetry, solution of the HFB equation is not yet implemented.

### 3.3 Configurations

**Keyword:** `PHNONE_NEU`

1, 00, 00 = `NUPAHO,KPNONE,KHNONE`

Neutron particle-hole excitations pertaining to the situation when all neutrons are in one common block (no simplex, signature, or parity is conserved). `NUPAHO` is the consecutive number from 1 to 5 (up to five sets of excitations can be specified in separate items). Particles are removed from the `KHNONE`-th state and put in the `KPNONE`-th state. At every stage of constructing excitations the Pauli exclusion principle has to be respected (particle removed from an occupied state and put in an empty state). Values equal zero have no effect. Note that for all neutrons sitting in one common block the reference configuration from which the particle-hole excitations are counted is defined by the total number of neutrons.

**Keyword:** `PHNONE_PRO`

1, 00, 00 = `NUPAHO,KPNONE,KHNONE`

Same as above but for the proton particle-hole excitations.

**Keyword:** `VACPAR_NEU`

44, 42 = `KVASIQ(0),KVASIQ(1)`

Numbers of lowest neutron states occupied in the two blocks, denoted by (+1) and (−1), of given parities, $\pi$=+1 and $\pi$=−1, respectively. These numbers define the parity reference configuration from which the particle-hole excitations are counted. The definitions of parity reference configuration and excitations are ignored unless `IPARTY`=1, or `IPARTY`=−1 and `ISIMPY`=`ISIGNY`=1. They are also ignored for `IPAIRI`=0.

**Keyword:** `VACPAR_PRO`

32, 34 = `KVASIQ(0),KVASIQ(1)`

Same as above but for the numbers of proton states.

**Keyword:** `PHPARI_NEU`

1, 00, 00, 00, 00 = `NUPAHO,KPSIQP,KPSIQM,KHSIQP,KHSIQM`

Neutron particle-hole excitations in the parity blocks. `NUPAHO` is the consecutive number from 1 to 5 (up to five sets of excitations can be specified in separate items). Particles are removed from the `KHSIQP`-th state in the (+1) block and from the `KHSIQM`-th state in the (−1) block, and put in the `KPSIQP`-th state in the (+1) block and in the `KPSIQM`-th state in the (−1) block. At every stage of constructing excitations the Pauli exclusion principle has to be respected (particle removed from an occupied state and put in an empty state). Values equal zero have no effect.

**Keyword:** `PHPARI_PRO`

1, 00, 00, 00, 00 = `NUPAHO,KPSIQP,KPSIQM,KHSIQP,KHSIQM`

Same as above but for the proton particle-hole excitations.

**Keyword:** `DIAPAR_NEU`

2, 2, 1, 1, 0, 0 = `KPFLIQ(0,0),KPFLIQ(1,0),`
`KHFLIQ(0,0),KHFLIQ(1,0),`
`KOFLIQ(0,0),KOFLIQ(1,0)`

Diabatic blocking of neutron single-particle parity configurations. Matrices `KPFLIQ` contain the indices of the particle states in the two parity blocks denoted by (+1) and (−1), of given parities, i.e., $\pi$=+1 and −1, respectively. Matrices `KHFLIQ` contain analogous indices of the hole states, and matrices `KOFLIQ` define type of blocking according to the following table:

| | | |
|---|---|---|
| KOFLIQ=0 | $\Longleftrightarrow$ | No diabatic blocking in the given parity block. |
| KOFLIQ=+1 | $\Longleftrightarrow$ | The state which has the *larger* $y$-alignment is occupied. |
| KOFLIQ=−1 | $\Longleftrightarrow$ | The state which has the *smaller* $y$-alignment is occupied. |
| KOFLIQ=+2 | $\Longleftrightarrow$ | The state which has the *larger* $y$-intrinsic spin is occupied. |
| KOFLIQ=−2 | $\Longleftrightarrow$ | The state which has the *smaller* $y$-intrinsic spin is occupied. |
| KOFLIQ=+3 | $\Longleftrightarrow$ | The state which has the *larger* $x$-alignment is occupied. |
| KOFLIQ=−3 | $\Longleftrightarrow$ | The state which has the *smaller* $x$-alignment is occupied. |
| KOFLIQ=+4 | $\Longleftrightarrow$ | The state which has the *larger* $x$-intrinsic spin is occupied. |
| KOFLIQ=−4 | $\Longleftrightarrow$ | The state which has the *smaller* $x$-intrinsic spin is occupied. |
| KOFLIQ=+5 | $\Longleftrightarrow$ | The state which has the *larger* $z$-alignment is occupied. |
| KOFLIQ=−5 | $\Longleftrightarrow$ | The state which has the *smaller* $z$-alignment is occupied. |
| KOFLIQ=+6 | $\Longleftrightarrow$ | The state which has the *larger* $z$-intrinsic spin is occupied. |
| KOFLIQ=−6 | $\Longleftrightarrow$ | The state which has the *smaller* $z$-intrinsic spin is occupied. |
| KOFLIQ=+7 | $\Longleftrightarrow$ | The state which has the *larger* $o$-alignment is occupied. |
| KOFLIQ=−7 | $\Longleftrightarrow$ | The state which has the *smaller* $o$-alignment is occupied. |
| KOFLIQ=+8 | $\Longleftrightarrow$ | The state which has the *larger* $o$-intrinsic spin is occupied. |
| KOFLIQ=−8 | $\Longleftrightarrow$ | The state which has the *smaller* $o$-intrinsic spin is occupied. |
| KOFLIQ=+9 | $\Longleftrightarrow$ | The state which has the *larger* multipole moment $Q_{20}$ is occupied. |
| KOFLIQ=−9 | $\Longleftrightarrow$ | The state which has the *smaller* multipole moment $Q_{20}$ is occupied. |
| KOFLIQ=+10 | $\Longleftrightarrow$ | The state which has the *larger* multipole moment $Q_{22}$ is occupied. |
| KOFLIQ=−10 | $\Longleftrightarrow$ | The state which has the *smaller* multipole moment $Q_{22}$ is occupied. |

Here, the $x$-, $y$-, and $z$-alignments or intrinsic spins denote projections of the total angular momentum or spin, respectively, on the $x$, $y$, and $z$ axes. Similarly, the $o$-alignment or $o$-intrinsic spin denotes analogous projections on the direction of the angular velocity $\boldsymbol{\omega}_J$. In version (v2.07f), observables KOFLIQ=$\pm3,\ldots,\pm10$ have been added to the above list. Identical extension of the list was also implemented for the diabatic blocking in the simplex or parity/signature configurations, see Sec. III-3.2.

Within the diabatic blocking procedure one does not predefine whether the particle or the hole state is occupied (like is the case when the particle-hole excitations are defined, see Section 3.4 of II). In each iteration the code calculates the average alignments (or average intrinsic spins, or average quadrupole moments) of both states (those defined by KPFLIM and KHFLIM), and occupies that state for which a larger, or a smaller value is obtained. Therefore, the order of both states in the Routhian spectrum is irrelevant.

The user is responsible for choosing the particle-state indices (in KPFLIM) only among those corresponding to empty single-particle states, and the hole-state indices (in KHFLIM) only among those corresponding to occupied single-particle states, see Section 3.4 of II.

**Keyword:** DIAPAR_PRO

2, 2, 1, 1, 0, 0 = KPFLIQ(0,1),KPFLIQ(1,1),
KHFLIQ(0,1),KHFLIQ(1,1),
KOFLIQ(0,1),KOFLIQ(1,1)

Same as above but for the diabatic blocking of proton single-particle parity configurations.

**Keyword:** DIANON_NEU

2, 1, 0 = KPFLIZ,KHFLIZ,KOFLIZ

Same as above but for the diabatic blocking of neutron single-particle configurations in the situation when all neutrons are in one common block.

**Keyword:** `DIANON_PRO`

2, 1, 0 = `KPFLIZ,KHFLIZ,KOFLIZ`

Same as above but for the diabatic blocking of proton single-particle configurations in the situation when all protons are in one common block.

### 3.4 Ensemble of specific parameters referred to as "Numerical data"

**Keyword:** `MAX_SURFAC`

0, 0 = `NSICON,NSIPRI`

Maximum multipolarities $\lambda$ of surface multipole moments, see Sec. 2.4, used in the code for the constraints, Eq. (4), and printed on the output, respectively. Values not larger than $\lambda$=7 are currently allowed.

**Keyword:** `MAX_MAGNET`

0, 0 = `NMACON,NMAPRI`

Same as above but for the magnetic moments. In version (v2.07f) constraints on magnetic moments are not yet implemented, which requires `NMACON`=0. Values not larger than $\lambda$=9 are currently allowed.

**Keyword:** `SLOW-PAIR`

0.5 = `SLOWPA`

Similarly as in the particle-hole channel, the prescription to calculate the pairing potential in the next iteration is to mix a given fraction $\epsilon$ of the pairing potential from the previous iteration, with the fraction $1-\epsilon$ of potential given by expression (15). `SLOWPA` gives the value of $\epsilon$ in the pairing channel, in analogy to `SLOWEV` and `SLOWOD`, Sec. II-3.5, which have been introduced for the particle-hole channel.

### 3.5 Constraints

**Keyword:** `OMISOY`

0.00 = `OMISOY`

Isovector angular frequency $\omega_{J1y}$ in the $y$ direction, see Sec. 2.3 and Eq. (1). A non-zero value of `OMISOY` requires `IROTAT`=1 and cannot be used for the BCS pairing, i.e., for `IPAIRI`=1 and `IPAHFB`=0.

**Keyword:** `OMEGA_XYZ`

0.00, 0.00, 0.00, 0 = `OMEGAX,OMEGAY,OMEGAZ,ITILAX`

For `ITILAX`=1, values of the three Cartesian components of the isoscalar angular frequency vector $\boldsymbol{\omega}_{J0}$, see Sec. 2.3 and Eq. (1). For `ITILAX`=0 these values are ignored. `ITILAX`=1 requires `IROTAT`=1, `ISIMPY`=0, and `IPAIRI`=0. A non-zero value of `OMEGAX`, `OMEGAY`, or `OMEGAZ` requires broken symmetry `ISIMTX`=0, `ISIMTY`=0, or `ISIMTZ`=0, respectively.

**Keyword:** `OMEGA_RTP`

0.00, 0.00, 0.00, 0 = `OMERAD,OMETHE,OMEPHI,ITILAX`

Same as above but for values of the standard spherical components of the isoscalar angular frequency, i.e., $\omega_{J0r}$, $\omega_{J0\theta}$, and $\omega_{J0\phi}$.

**Keyword:** `OMISO_XYZ`

0.00, 0.00, 0.00, 0 = `OMISOX,OMISOY,OMISOZ,ITISAX`

For `ITISAX`=1, values of the three Cartesian components of the isovector angular frequency vector $\boldsymbol{\omega}_{J1}$, see Sec. 2.3 and Eq. (1). For `ITISAX`=0 these values are ignored. `ITISAX`=1 requires `IROTAT`=1, `ISIMPY`=0, and `IPAIRI`=0. A non-zero value of `OMISOX`, `OMISOY`, or `OMISOZ` requires broken symmetry `ISIMTX`=0, `ISIMTY`=0, or `ISIMTZ`=0, respectively.

**Keyword:** `OMEGA_TURN`

0 = `IMOVAX`

For `IMOVAX`=1 or −1, the isoscalar angular frequency vector $\boldsymbol{\omega}_{J0}$, see Sec. 2.3, is in each iteration set in the direction of the total angular momentum $\boldsymbol{J}_0$, or opposite to this direction, respectively, while its length is kept equal to that fixed by the sum of squares of `OMEGAX`, `OMEGAY`, and `OMEGAZ`. For `IMOVAX`=0, vector $\boldsymbol{\omega}_{J0}$ is not changed during the iteration. `IMOVAX`=1 or −1 requires `ITILAX`=1, `ITISAX`=0, and `IFLAGA`=0.

**Keyword:** `SURFCONSTR`

2, 0, 0.0, 0.0, 0 = `LAMBDA`, `MIU`, `STIFFS`, `SASKED`, `IFLAGS`

For `IFLAGS`=1, the surface mass multipole moment of the given multipolarity $\lambda$ and $\mu$ is constrained. Values of `LAMBDA`, `MIU`, `STIFFS`, and `SASKED` correspond respectively to $\lambda$, $\mu$, $C^S_{\lambda\mu}$, and $\bar{Q}^S_{\lambda\mu}$ in Eq. (4). For `IFLAGS`=0, there is no constraint in the given multipolarity.

**Keyword:** `SPICON_XYZ`

0.0, 0.0, 0 = `STIFFI(1),ASKEDI(1),IFLAGI(1)`
0.0, 0.0, 0 = `STIFFI(2),ASKEDI(2),IFLAGI(2)`
0.0, 0.0, 0 = `STIFFI(3),ASKEDI(3),IFLAGI(3)`

For `IFLAGI`=1, the quadratic constraint on one of the Cartesian components of angular momentum is used together with the linear constraint, see Sec. 2.3 and Eq. (1). Values of `STIFFI` and `ASKEDI` correspond respectively to $C_{Ja}$ and $\bar{J}_{J0a}$ in Eq. (1), where $a$=$x$, $y$, or $z$. For `IFLAGI`=0, there is no quadratic constraint on a given component. `IFLAGI(1)`=1, `IFLAGI(2)`=1, or `IFLAGI(3)`=1 requires broken symmetry `ISIMTX`=0, `ISIMTY`=0, or `ISIMTZ`=0, respectively.

**Keyword:** `SPICON_OME`

0.0, 0.0, 0 = `STIFFA,ASKEDA,IFLAGA`

For `IFLAGA`=1, the quadratic constraint on the angle between the angular frequency and angular momentum vectors, see Sec. 2.3 and Eq. (1). In version (v2.07f) the angle is constrained to zero. Value of `STIFFA` corresponds to $C_A$ in Eq. (1). Value of `ASKEDA` must be set to 0; this variable is reserved for a future implementation of the constraint to a non-zero angle. For `IFLAGA`=0, there is no quadratic constraint on the angle. `IFLAGA`=1 requires `ISIMPY`=0, `IROTAT`=1, `ITILAX`=1, `ITISAX`=0, `IFLAGI(1)`=0, `IFLAGI(2)`=0, `IFLAGI(3)`=0, `ISIMTX`=0, `ISIMTY`=0, `ISIMTZ`=0, and a non-zero value of sum of squares of `OMEGAX`, `OMEGAY`, and `OMEGAZ`.

### 3.6 Output-file parameters

**Keyword:** `PRINT-MOME`

1, 1, 1 = `IPRI_N,IPRI_P,IPRI_T`

The code prints or not values of the neutron, proton, and total moments (multipole, surface multipole, or magnetic) only if the corresponding parameters IPRI_N, IPRI_P, and IPRI_T equal 1 or 0, respectively.

**Keyword:** PRINT-INTR

1 = INTRIP

The code prints or not the intrinsic-frame values of moments (multipole, surface multipole, or magnetic) and angular momenta only if INTRIP equals 1 or 0, respectively.

**Keyword:** EQUASI_MAX

10.0 = EMAXQU

The code prints tables of quasiparticle properties for the HFB states with values of the quasiparticle energies smaller than EMAXQU. No table is printed unless EMAXQU>0.

### 3.7 Files

**Keyword:** RECORDSAVE

1 = IWRIRE

For IWRIRE=1, the record file, see Sec. II-3.9, is saved on disc after each iteration is completed. For IWRIRE=0, it is saved only once, after all iterations are completed. This option is useful on systems where the speed of writing large amounts of data to disc could hamper the system performance. For IWRIRE=−1, the record file is never saved, and the run cannot be later continued.

### 3.8 Starting the iteration

**Keyword:** CONT_PAIRI

0 = IPCONT

For IPCONT=1, results stored in the replay file will be used to define pairing properties in the first iteration; otherwise values read in matrices FERINI and DELINI (see Sec. 3.1) will be used. When pairing correlations are taken into account (i.e., for IPAIRI=1, see Sec. II-3.3), and if a smooth restart and continuation of iterations from previously stored results is required, value of IPCONT=1 must be used. IPCONT=1 is incompatible with either of IPAIRI=0 or ICONTI=0.

**Keyword:** CONT_OMEGA

0 = IOCONT

For IOCONT=1, the angular momentum vector $\boldsymbol{J}_0$ stored in the replay file will be used to define direction of the angular frequency vector $\boldsymbol{\omega}_{J0}$, see Sec. 2.3 in the first iteration; otherwise values read in OMEGAX, OMEGAY, and OMEGAZ (see Sec. 3.5) will be used. IOCONT=1 requires IMOVAX=1 or −1.

## 4 Output file

Together with the FORTRAN source code in the file hfodd.f, an example of the output file is provided in la132-a.out. Selected lines from this file are presented in section TEST RUN OUTPUT below. This output file corresponds to the input file la132-a.dat reproduced in section TEST RUN INPUT below.

The output file contains two consecutive runs of the code HFODD. The first run starts from the Nilsson model, see Sec. I-3.10, and in 95 iterations finds a triaxial solution in $^{132}$La for $\hbar\boldsymbol{\omega}$=0. The triaxial shape is forced by using appropriate constraints for moments $\langle\hat{Q}_{20}\rangle$ and $\langle\hat{Q}_{22}\rangle$. Such a solution is obtained with parity and signature symmetries conserved, and with the time-reversal broken in order to describe odd numbers of neutrons and protons. Therefore, even for $\hbar\boldsymbol{\omega}$=0 one obtains solution with a non-zero value of the angular momentum, which is oriented along the $y$ axis. Since this first run is performed within conditions that have already been available in the previous version (v1.75r), in section TEST RUN OUTPUT we skip the corresponding output lines.

The second run starts from the solution obtained in the first run and performs calculations with all symmetries broken except parity, and for the initial angular momentum vector of $\hbar\omega_x = \hbar\omega_y = \hbar\omega_z = (0.25/\sqrt{3})$ MeV. Since for such a low value of the angular frequency, only a planar solution exists [20], the nucleus has to turn in space in such way that the angular momentum vector coincides with the symmetry plane spanned by the short and long axes of the mass distribution. Standard iteration, with `IMOVAX`=0, see Secs. 2.3 and 3.5, requires several thousand iterations to achieve the alignment of the angular momentum and angular frequency vectors below the angle of 0.01°. On the other hand, iteration with `IMOVAX`=1, as shown in the example of file `la132-a.dat`, achieves the alignment of 0.000001° within 92 iterations.

Section `SKYRME FORCE DEFINITION` lists the name and parameters of the Skyrme force together with parameters `KETA_J`, `KETA_W`, `KETACM`, and `KETA_M` that define the way the given force should be used (see Sec. 3.1).

Section `CALCULATIONS WITH THE TILTED-AXIS CRANKING` gives values of components of the angular frequency vector and its length. For switches `IMOVAX`=1 or −1 and `IOCONT`=0, these vales are used only in the first iteration, and later ignored, because the angular frequency vector is in each iteration readjusted to be aligned or anti-aligned with the angular momentum vector.

Section `PARITY CONFIGURATIONS` gives numbers of neutron and proton states in the two parity blocks, see Sec. 3.3.

Section `CONVERGENCE REPORT` gives information on the performed iterations. One line per one iteration lists values described in Sec. II-4, except that version (v2.07f) prints the multipole moments in the intrinsic frame, the last before last column gives the angle between the angular frequency and angular momentum vectors, and the last column gives the total pairing energy.

Section `SINGLE-PARTICLE PROPERTIES` gives information on the s.p. states. For broken simplex symmetry, three projections of the total angular momentum and intrinsic spin are printed in the first and second line, respectively, for each s.p. state.

Section `EULER ANGLES OF THE PRINCIPAL-AXES FRAME` gives the standard Euler angles $\alpha$, $\beta$, and $\gamma$ [8] in degrees, see Sec. 2.4, which define the orientation of the principal-axes (intrinsic) frame of reference with respect to the original frame.

Section `MULTIPOLE MOMENTS [UNITS: (10 FERMI)^LAMBDA]` gives values of multipole moments with respect to the original frame of reference. Similarly, section `MULTIPOLE MOMENTS [UNITS: (10 FERMI)^LAMBDA] [INTRINSIC FRAME]` gives analogous values with respect to the intrinsic frame of reference. Whenever the parity symmetry is broken, yet another analogous section gives information on values of multipole moments with respect to the center-of-mass reference frame.

Sections `MAGNETIC MOMENTS [MAGNETON*FERMI^(LAMBDA-1)]` and `MAGNETIC MOMENTS`

[MAGNETON*FERMI^(LAMBDA-1)] [INTRINSIC FRAME] give values of magnetic moments with respect to the original and intrinsic reference frame, respectively.

Section ANGULAR MOMENTA IN THE THREE CARTESIAN DIRECTIONS gives values of neutron, proton, and total projections of the total angular momentum and intrinsic spin on the three Cartesian axes.

Sections NEUTRON CONFIGURATIONS and PROTON CONFIGURATIONS give visual representation of states occupied in the parity blocks, in analogy to those described in Sec. I-4.

Section ENERGIES gives summary of energies calculated in the last iteration. Apart from entries described in Sec. I-4, it gives in addition the pairing rearrangement energies P-REARR, pairing gaps PAIRGAP, Fermi energies E-FERMI, and surface multipole constraint energies CONSTR. (SURF).

Alternatively, one can perform the second run with the $x$-simplex$^T$ conserved (ISIMTX=1) and the initial angular frequency vector of $\hbar\omega_y = \hbar\omega_z = (0.25/\sqrt{2})$ MeV and $\hbar\omega_x$=0. In this way, the angular momentum is confined within the symmetry plane $y$-$z$ spanned by the short and long axes of the mass distribution. The corresponding input and output data files are provided in files la132-b.dat and la132-b.out, respectively. Since the planar solution in question corresponds to projections of the angular momentum on the short and long axes that are not equal to one another, the nucleus has now turn in space about the $x$ axis. For IMOVAX=0, this process is also very slow, while for IMOVAX=1, as in file la132-b.dat it takes again only 92 iterations. Apart from different orientations in space, solutions obtained by using input data files la132-a.dat and la132-b.dat are entirely equivalent. Output file la132-b.out is not shown in section TEST RUN OUTPUT below.

# 5 FORTRAN source file

The FORTRAN source code in is provided in the file hfodd.f and can be modified in several places, as described in this section.

## 5.1 Library subroutines

The code HFODD requires an external subroutine which diagonalizes complex hermitian matrices. Version (v1.60r), see II, has been prepared with an interface to the NAGLIB subroutine F02AXE, and version (v1.75r), see III, with an interface to the LAPACK subroutine ZHPEV, which can be downloaded (with dependencies) from
http://netlib2.cs.utk.edu/cgi-bin/netlibfiles.pl?filename=/lapack/complex16/zhpev.f
In the present version (v2.07f) both these interfaces remain supported, and can be activated as described in II and III, respectively.

In addition, in version (v2.07f) we have implemented interface to the LAPACK subroutine ZHPEVX, which can be downloaded (with dependencies) from
http://netlib2.cs.utk.edu/cgi-bin/netlibfiles.pl?filename=/lapack/complex16/zhpevx.f
This subroutine finds not all, but only the lowest eigenvectors, and hence performs calculations in less CPU time; the gain is particularly significant for large HO bases. Numbers of eigenvectors to be found are defined by the size of the HO phasespace, see keyword PHASESPACE in Sec. III-3.1.

Subroutine ZHPEVX and its dependencies are in the `REAL*8/COMPLEX*16` version, and should be compiled without promoting real numbers to the double precision. On the other hand, the code HFODD itself does require compilation with an option promoting to double precision. Therefore, the code and the ZHPEVX package should be compiled separately, and then should be linked together.

In order to activate the interface to the LAPACK ZHPEVX subroutine, the following modifications of the code HFODD (v2.07f) have to be made:

1. Change everywhere the value of parameter `I_CRAY=1` into `I_CRAY=0`.

2. Change everywhere the value of parameter `IZHPEV=0` into `IZHPEV=2`.

3. If your compiler does not support undefined externals, or subroutines called with different parameters, remove calls to subroutines CGEMM, F02AXE, and ZHPEV.

### 5.2 FORTRAN-90 version

The code HFODD version (v2.07f) is written in FORTRAN-77. However, on several platforms several tested FORTRAN-77 compilers did not perform well with respect to the memory management. In particular, they often distributed large matrices on computer's stack and heap in a wasteful way, and then the available memory could be insufficient for the code to execute, especially for large HO bases. Therefore, crucial parts of the code have been promoted to FORTRAN-90, with an active use of explicit memory allocation and deallocation. In the FORTRAN source code provided in the file `hfodd.f`, all these FORTRAN-90 features are commented out and inactive. However, very simple modifications of the source code can easily be performed to transform code HFODD to FORTRAN-90. To this end, the user should run an automated editor script which will perform the following two operations:

1. Replace all strings "C F90" by spaces, and

2. Remove two lines of code (or three lines of the file) after lines marked by the string: "C IN F90 VERSION PLEASE REMOVE THE TWO FOLLOWING LINES".

The FORTRAN-90 version should be compiled with the "fixed form" non-standard option of the FORTRAN-90 compiler.

### 5.3 Printing the execution times

After successful completion of execution, the code HFODD prints the table of times spent in principal subroutines, as well as the total execution time. Since implementation of the time-checking functions widely varies between compilers and platforms, in the file `hfodd.f` all references to these functions are commented out and inactive. However, very simple modifications of the source code can easily be performed to activate one of the following options:

1. Replace all strings "CCPUT" by spaces in order to activate a call to the FORTRAN-90 subroutine CPU_TIME, which returns CPU execution times.

2. Replace all strings "CETIM" by spaces in order to activate a call to function ETIME, which returns CPU execution times.

3. Replace all strings "CUNIX" by spaces in order to activate a call to the UNIX function TIME, which returns wall clock execution times.

4. Replace all strings "CSAL" by spaces in order to activate a call to the SALFORD© FORTRAN subroutine DCLOCK@, which returns wall clock execution times.

5. Replace all strings "CF32" by spaces in order to activate a call to the MICROSOFT© F32 subroutine GETTIM, which returns wall clock execution times.

Other time-checking functions, available to the user, can be analogously called in the subroutine CPUTIM.

# 6 Acknowledgments

The authors would like to thank Jerzy Dudek for his hospitality during their stay at the *Institut de Recherches Subatomiques* in Strasbourg, where part of this work was done. This research was supported in part by the Polish Committee for Scientific Research (KBN) under Contract No. 5 P03B 014 21, by the Foundation for Polish Science (FNP), by the French-Polish integrated actions programme POLONIUM, by the U.S. Department of Energy under Contract Nos. DE-FG02-96ER40963 (University of Tennessee), and DE-AC05-00OR22725 with UT-Battelle, LLC (Oak Ridge National Laboratory), and by the computational grant from the Interdisciplinary Centre for Mathematical and Computational Modeling (ICM) of the Warsaw University.

## TEST RUN INPUT

```
=============================================================================
| This file (la132-a.dat) contains the input data for the code HFODD.       |
| It defines two consecutive runs that lead to a planar solution in 132La. |
=============================================================================
                        ------------  Nuclide  -------------
NUCLIDE
            75 57
                        ----------  Convergence  -----------
ITERATIONS
            5000
ITERAT_EPS
            0.000001
MAXANTIOSC
            5
                        -------  Skyrme interaction  -------
SKYRME-STD
            1 0 0 0 0
                        ---------  Configurations  ---------
VACSIG_NEU       PPSP  PPSM  PMSP  PMSM
                  19    19    19    18
VACSIG_PRO       PPSP  PPSM  PMSP  PMSM
                  15    15    14    13
                        ------------  HO basis  ------------
BASIS_SIZE
            10 286 800.
SURFAC_PAR
            75 57 1.230
SURFAC_DEF
            2 0 0.000
SURFAC_DEF
            4 0 0.000
                        ----------  Output data  -----------
MAX_MAGNET
            0 3
                        -----------  File names  -----------
REVIEWFILE
            la132-a.rev
REPLAYFILE
            la132-a.rec
RECORDFILE
            la132-a.rec
                        ----------  Constraints  -----------
OMEGAY
            0.000
MULTCONSTR
            2 0 0.5  7.0    1
MULTCONSTR
            2 2 0.5  7.0    1
                        ---------  Starting point  ---------
RESTART
            0
                        ------------  Calculate  -----------
EXECUTE
                        -----------  Symmetries  -----------
SIMPLEXY
            0
SIGNATUREY
            0
PARITY
            1
TSIMPLEX3D
            0 0 0
                        ---------  Configurations  ---------
VACPAR_NEU             PARP  PARM
                        38    37
VACPAR_PRO             PARP  PARM
                        30    27
                        ----------  Constraints  -----------
OMEGA_XYZ
            0.1443375673 0.1443375673 0.1443375673 1
OMEGA_TURN
            1
MULTCONSTR
            2 0 0.5  7.0    0
MULTCONSTR
            2 2 0.5  7.0    0
                        ---------  Starting point  ---------
RESTART
            1
CONT_OMEGA
            0
                        ------------  Calculate  -----------
EXECUTE
                        ------------  Terminate  -----------
ALL_DONE
```

## TEST RUN OUTPUT

```
*******************************************************************************
*                                                                             *
*    HFODD    HFODD    HFODD    HFODD    HFODD    HFODD    HFODD    HFODD     *
*                                                                             *
*******************************************************************************
*                                                                             *
*              SKYRME-HARTREE-FOCK-BOGOLYUBOV CODE VERSION: 2.07F             *
*                                                                             *
*               NO SYMMETRY-PLANES AND NO TIME-REVERSAL SYMMETRY              *
*                                                                             *
*                 DEFORMED CARTESIAN HARMONIC-OSCILLATOR BASIS                *
*                                                                             *
*******************************************************************************
*                                                                             *
*                 J. DOBACZEWSKI, J. DUDEK, AND P. OLBRATOWSKI                *
*                                                                             *
*                INSTITUT DE RECHERCHES SUBATOMIQUES, STRASBOURG              *
*                                                                             *
*                    INSTYTUT FIZYKI TEORETYCZNEJ, WARSZAWA                   *
*                                                                             *
*                                   1993-2004                                 *
*                                                                             *
*******************************************************************************
*******************************************************************************
*                                                                             *
*                              SKYRME FORCE DEFINITION                        *
*                                                                             *
*******************************************************************************
*                                                                             *
* PARAMETER SET SKM*:  T0= -2645.00  T1=   410.00  T2=  -135.00  T3= 15595.00 *
*                                                                             *
* POWER=1/6 W=130.000  X0=  0.09000  X1=  0.00000  X2=  0.00000  X3=  0.00000 *
*                                                                             *
* ETA_J=0 ETA_W=0 ETACM=0 ETA_M=1   HBM= 20.73000  (SKYRME-FORCE  SPIN-ORBIT) *
*                                                                             *
*******************************************************************************
*******************************************************************************
*                                                                             *
* CALCULATIONS WITH ADJUSTABLE ISOSCALAR ROTATIONAL FREQUENCY VECTOR,  WHICH  *
* IN EACH ITERATION IS TURNED TO BE PARALLEL TO THE ANGULAR MOMENTUM  VECTOR  *
*                                                                             *
* DIRECTION OF THE ROTATIONAL FREQUENCY VECTOR THAT IS GIVEN BELOW  WILL  BE  *
* USED ONLY IN THE FIRST ITERATION, AND LATER WILL BE IGNORED                 *
*                                                                             *
*******************************************************************************
*                                                                             *
* CALCULATIONS WITH THE TILTED-AXIS CRANKING                                  *
*                                                                             *
* FOR THE ISOSCALAR ROTATIONAL FREQUENCIES OF:                                 *
*                                                                             *
* OMEGAX = 0.144338  OMEGAY = 0.144338  OMEGAZ = 0.144338   OMEGA = 0.250000  *
*                                                                             *
*******************************************************************************
*                                                                             *
*  PARITY CONFIGURATIONS:                                                     *
*                                                                             *
*               V A C U U M       P A R T I C L E S        H O L E S          *
*               ===========       =================       ===========         *
*               ( + ) ( - )          ( + ) ( - )          ( + ) ( - )         *
*                                                                             *
*  NEUTRONS:      38    37              0     0              0     0          *
*  PROTONS :      30    27              0     0              0     0          *
*                                                                             *
*******************************************************************************
*******************************************************************************
*                                                                             *
* CONVERGENCE REPORT                                                          *
*                                                                             *
*******************************************************************************
*                                                                             *
* ITER     ENERGY     STABILITY    Q20    GAMMA   SPIN  OMEGA   ANGLE   EPAIR *
*                                                                             *
*  96  -1093.673718  -0.232831   6.890   45.36  11.28  0.250  13.308    0.00 *
*  97  -1093.850272   0.025915   6.876   45.40  11.40  0.250   8.286    0.00 *
*  98  -1093.854682   0.044356   6.863   45.43  11.59  0.250   4.645    0.00 *
*  99  -1093.844776   0.041488   6.852   45.46  11.71  0.250   2.825    0.00 *
*  100 -1093.832812   0.036000   6.843   45.49  11.79  0.250   1.773    0.00 *
*  101 -1093.823120   0.030728   6.834   45.51  11.85  0.250   1.207    0.00 *
*  102 -1093.815613   0.026215   6.827   45.52  11.89  0.250   0.857    0.00 *
*  103 -1093.810028   0.022473   6.821   45.54  11.92  0.250   0.637    0.00 *
*  179 -1093.792431   1.31E-06   6.780   45.62  12.00  0.250 1.7E-06    0.00 *
*  180 -1093.792431   1.15E-06   6.780   45.62  12.00  0.250 1.5E-06    0.00 *
*  181 -1093.792431   1.01E-06   6.780   45.62  12.00  0.250 8.5E-07    0.00 *
*  182 -1093.792431   8.94E-07   6.780   45.62  12.00  0.250 1.5E-06    0.00 *
*  183 -1093.792431   7.88E-07   6.780   45.62  12.00  0.250 8.5E-07    0.00 *
*  184 -1093.792431   6.95E-07   6.780   45.62  12.00  0.250 1.5E-06    0.00 *
*  185 -1093.792431   6.13E-07   6.780   45.62  12.00  0.250 8.5E-07    0.00 *
*  186 -1093.792431   5.40E-07   6.780   45.62  12.00  0.250   0.000    0.00 *
*                                                                             *
*******************************************************************************
```

```
*******************************************************************************
*                                                                             *
*  SINGLE-PARTICLE PROPERTIES: HARTREE-FOCK                        NEUTRONS   *
*                                                                             *
*******************************************************************************
*                                                                             *
*  NO)   ENERGY   (P=+,P=-)   | N,nz,lz,OMEG>  <P>    <JX>    <JY>    <JZ>    *
*                                                                             *
*  71)  -10.460   (  0, 34)   | 5, 1, 4, 9/2> -100   0.029  -0.189   2.078    *
*                                                    0.003  -0.022   0.223    *
*  72)  -10.338   ( 38,  0)   | 4, 0, 2, 3/2>  100  -0.007  -0.811  -0.436    *
*                                                    0.005   0.036   0.372    *
*  73)  -10.042   (  0, 35)   | 5, 2, 3, 7/2> -100  -0.034  -0.579  -2.392    *
*                                                   -0.004  -0.027  -0.285    *
*  74)   -9.318   (  0, 36)   | 5, 0, 5,11/2> -100   0.070  -0.630   5.021    *
*                                                    0.007  -0.034   0.495    *
*  75)   -9.145   (  0, 37)   | 5, 1, 4, 9/2> -100  -0.053  -0.580  -3.765    *
*                                                   -0.006  -0.040  -0.429    *
*  76)   -8.966   ( 39,  0)   | 4, 1, 1, 1/2>  100   0.005   0.589   0.287    *
*                                                   -0.003  -0.011  -0.240    *
*  77)   -8.637   ( 40,  0)   | 4, 0, 2, 5/2>  100  -0.004  -0.152  -0.282    *
*                                                   -0.002   0.045  -0.168    *
*  78)   -8.315   ( 41,  0)   | 4, 1, 1, 1/2>  100  -0.003   0.295  -0.247    *
*                                                    0.001   0.013   0.065    *
*  79)   -8.018   (  0, 38)   | 5, 2, 1, 1/2> -100   0.006   3.305   0.150    *
*                                                    0.001   0.332   0.034    *
*  80)   -7.731   (  0, 39)   | 5, 0, 5,11/2> -100  -0.068  -0.541  -4.847    *
*                                                   -0.007  -0.029  -0.496    *
*                                                                             *
*******************************************************************************
*******************************************************************************
*                                                                             *
*  EULER ANGLES OF THE PRINCIPAL-AXES FRAME IN DEGREES                 TOTAL  *
*                                                                             *
*  ALPHA = -1.65924  BETA =  0.80185  GAMMA =  1.72034                        *
*                                                                             *
*******************************************************************************

*******************************************************************************
*                                                                             *
*  MULTIPOLE MOMENTS [UNITS:  (10 FERMI)^LAMBDA]                       TOTAL  *
*                                                                             *
*  REAL PART  FOR A NON-NEGATIVE PROJECTION                                   *
*  IMAGINARY PART FOR A NEGATIVE PROJECTION                                   *
*                                                                             *
*******************************************************************************
*                                                                             *
*  Q00 =132.0000  .............  .............  .............  .............  *
*                 .............  .............  .............  .............  *
*                                                                             *
*  Q10 =    ZERO  Q1+1=    ZERO  .............  .............  .............  *
*                 Q1-1=    ZERO  .............  .............  .............  *
*                                                                             *
*  Q20 =  6.7791  Q2+1=  0.0194  Q2+2=  6.9293  .............  .............  *
*                 Q2-1= -0.0022  Q2-2= -0.0147  .............  .............  *
*                                                                             *
*  Q30 =    ZERO  Q3+1=    ZERO  Q3+2=    ZERO  Q3+3=    ZERO  .............  *
*                 Q3-1=    ZERO  Q3-2=    ZERO  Q3-3=    ZERO  .............  *
*                                                                             *
*  Q40 =  0.0849  Q4+1=  0.0039  Q4+2=  0.2199  Q4+3= -0.0038  Q4+4=  0.0994  *
*                 Q4-1= -0.0101  Q4-2=-7.9E-04  Q4-3= 7.9E-04  Q4-4=-4.1E-04  *
*                                                                             *
*******************************************************************************
*******************************************************************************
*                                                                             *
*  MAGNETIC MOMENTS  [MAGNETON*FERMI^(LAMBDA-1)]                       TOTAL  *
*                                                                             *
*  REAL PART  FOR A NON-NEGATIVE PROJECTION                                   *
*  IMAGINARY PART FOR A NEGATIVE PROJECTION                                   *
*                                                                             *
*******************************************************************************
*                                                                             *
*  M10 =  0.5491  M1+1= -0.0083  .............  .............  .............  *
*                 M1-1=  2.7154  .............  .............  .............  *
*                                                                             *
*  M20 =    ZERO  M2+1=    ZERO  M2+2=    ZERO  .............  .............  *
*                 M2-1=    ZERO  M2-2=    ZERO  .............  .............  *
*                                                                             *
*  M30 = 54.9564  M3+1= -1.3218  M3+2=  3.5784  M3+3= -0.2678  .............  *
*                 M3-1= 64.2708  M3-2=  0.3102  M3-3= 64.4256  .............  *
*                                                                             *
*******************************************************************************
*                                                                             *
*  CARTESIAN COMPONENTS OF MAGNETIC DIPOLE MOMENT                      TOTAL  *
*                                                                             *
*  M1_X =   0.0241    M1_Y =   7.8594    M1_Z =   1.1237    M1 =     7.9394  *
*                                                                             *
*******************************************************************************
```

```
******************************************************************************
*                                                                            *
*  MULTIPOLE MOMENTS [UNITS:  (10 FERMI)^LAMBDA] [INTRINSIC FRAME]     TOTAL *
*                                                                            *
*  REAL PART  FOR A NON-NEGATIVE PROJECTION                                  *
*  IMAGINARY PART FOR A NEGATIVE PROJECTION                                  *
*                                                                            *
******************************************************************************
*                                                                            *
*  Q00 =132.0000  .............  .............  .............  ............. *
*                 .............  .............  .............  ............. *
*                                                                            *
*  Q10 =    ZERO  Q1+1=    ZERO  .............  .............  ............. *
*                 Q1-1=    ZERO  .............  .............  ............. *
*                                                                            *
*  Q20 =  6.7799  Q2+1=    ZERO  Q2+2=  6.9289  .............  ............. *
*                 Q2-1=    ZERO  Q2-2=    ZERO  .............  ............. *
*                                                                            *
*  Q30 =    ZERO  Q3+1=    ZERO  Q3+2=    ZERO  Q3+3=    ZERO  ............. *
*                 Q3-1=    ZERO  Q3-2=    ZERO  Q3-3=    ZERO  ............. *
*                                                                            *
*  Q40 =  0.0848  Q4+1= 9.3E-09  Q4+2=  0.2200  Q4+3=-6.3E-09  Q4+4=  0.0994 *
*                 Q4-1= -0.0103  Q4-2= 6.7E-09  Q4-3= 5.7E-04  Q4-4=-1.3E-08 *
*                                                                            *
******************************************************************************
******************************************************************************
*                                                                            *
*  MAGNETIC MOMENTS  [MAGNETON*FERMI^(LAMBDA-1)] [INTRINSIC FRAME]     TOTAL *
*                                                                            *
*  REAL PART  FOR A NON-NEGATIVE PROJECTION                                  *
*  IMAGINARY PART FOR A NEGATIVE PROJECTION                                  *
*                                                                            *
******************************************************************************
*                                                                            *
*  M10 =  0.5507  M1+1= 1.4E-07  .............  .............  ............. *
*                 M1-1=  2.7152  .............  .............  ............. *
*                                                                            *
*  M20 =    ZERO  M2+1=    ZERO  M2+2=    ZERO  .............  ............. *
*                 M2-1=    ZERO  M2-2=    ZERO  .............  ............. *
*                                                                            *
*  M30 = 55.0803  M3+1=-4.0E-06  M3+2=  3.6402  M3+3= 2.3E-06  ............. *
*                 M3-1= 64.2333  M3-2= 4.5E-07  M3-3= 64.4215  ............. *
*                                                                            *
******************************************************************************
*                                                                            *
*  CARTESIAN COMPONENTS OF MAGNETIC DIPOLE MOMENT [INTRINSIC FRAME]    TOTAL *
*                                                                            *
*  M1_X =   0.0000    M1_Y =   7.8590    M1_Z =   1.1271     M1 =     7.9394 *
*                                                                            *
******************************************************************************
******************************************************************************
*                                                                            *
*  ANGULAR MOMENTA IN THE THREE CARTESIAN DIRECTIONS                         *
*                                                                            *
******************************************************************************
*                                                                            *
*           SPIN IN X-DIRECTION   SPIN IN Y-DIRECTION   SPIN IN Z-DIRECTION  *
*           -------------------   -------------------   -------------------  *
*           INTRI  ORBIT  TOTAL   INTRI  ORBIT  TOTAL   INTRI  ORBIT  TOTAL  *
*                                                                            *
*  NEUTRONS 0.006  0.101  0.107   0.073  0.518  0.591   0.428  7.209  7.637  *
*   PROTONS 0.002  0.035  0.037   0.392  5.949  6.341   0.131  2.030  2.160  *
*  --------                                                                  *
*     TOTAL 0.008  0.136  0.144   0.466  6.466  6.932   0.559  9.238  9.797  *
*                                                                            *
******************************************************************************
*                                                                            *
*  ROTATIONAL FREQUENCIES IN THE THREE CARTESIAN DIRECTIONS:                 *
*                                                                            *
*  OMEGAX = 0.003009  OMEGAY = 0.144390  OMEGAZ = 0.204065  OMEGA = 0.250000 *
*                                                                            *
******************************************************************************
*                                                                            *
*  ANGLE BETWEEN ANGULAR MOMENTUM AND ANGULAR FREQUENCY VECTORS= 8.5E-07 DEG *
*                                                                            *
******************************************************************************
*                                                                            *
*  ANGULAR MOMENTA IN THE INTRINSIC REFERENCE FRAME                          *
*                                                                            *
******************************************************************************
*                                                                            *
*           SPIN IN X-DIRECTION   SPIN IN Y-DIRECTION   SPIN IN Z-DIRECTION  *
*           -------------------   -------------------   -------------------  *
*           INTRI  ORBIT  TOTAL   INTRI  ORBIT  TOTAL   INTRI  ORBIT  TOTAL  *
*                                                                            *
*  NEUTRONS 0.000  0.000  0.000   0.073  0.515  0.588   0.428  7.210  7.638  *
*   PROTONS 0.000  0.000  0.000   0.392  5.948  6.340   0.131  2.032  2.163  *
*  --------                                                                  *
*     TOTAL 0.000  0.000  0.000   0.466  6.463  6.928   0.559  9.242  9.801  *
*                                                                            *
******************************************************************************
```

```
*****************************************************************************
*                                                                           *
*                          NEUTRON  CONFIGURATIONS                          *
*                          =======================                          *
*      PAR  27 28 29 30 31 32 33 34 35 36 37 38 39 40 41 42 43 44 45 46 47  *
*      ---  -------------------------------------------------------------- *
*                                                                           *
*  CONF: (+)   1  1  1  1  1  1  1  1  1  1  1  1  0  0  0  0  0  0  0  0  0  *
*  VACC: (+)   1  1  1  1  1  1  1  1  1  1  1  1  0  0  0  0  0  0  0  0  0  *
*                                                                           *
*  CONF: (-)   1  1  1  1  1  1  1  1  1  1  1  0  0  0  0  0  0  0  0  0  0  *
*  VACC: (-)   1  1  1  1  1  1  1  1  1  1  1  0  0  0  0  0  0  0  0  0  0  *
*                                                                           *
*****************************************************************************
*                                                                           *
*                          PROTON   CONFIGURATIONS                          *
*                          ======================                           *
*      PAR  17 18 19 20 21 22 23 24 25 26 27 28 29 30 31 32 33 34 35 36 37  *
*      ---  -------------------------------------------------------------- *
*                                                                           *
*  CONF: (+)   1  1  1  1  1  1  1  1  1  1  1  1  1  1  0  0  0  0  0  0  0  *
*  VACC: (+)   1  1  1  1  1  1  1  1  1  1  1  1  1  1  0  0  0  0  0  0  0  *
*                                                                           *
*  CONF: (-)   1  1  1  1  1  1  1  1  1  1  1  0  0  0  0  0  0  0  0  0  0  *
*  VACC: (-)   1  1  1  1  1  1  1  1  1  1  1  0  0  0  0  0  0  0  0  0  0  *
*                                                                           *
*****************************************************************************
*                                                                           *
*                               ENERGIES (MEV)                              *
*                                                                           *
*****************************************************************************
*                                                                           *
*  KINETIC: (NEU)=  1460.491929    (PRO)=   958.114770    (TOT)=  2418.606700 *
*  SUM EPS: (NEU)= -1764.973911    (PRO)= -1047.797062    (TOT)= -2812.770974 *
*  PAIRING: (NEU)=     0.000000    (PRO)=     0.000000    (TOT)=     0.000000 *
*  P-REARR: (NEU)=     0.000000    (PRO)=     0.000000    (TOT)=     0.000000 *
*                                                                           *
*  PAIRGAP: (NEU)=     0.000000    (PRO)=     0.000000    THIS SPACE IS EMPTY *
*  E-FERMI: (NEU)=    -9.055110    (PRO)=    -4.947826    AND AWAITS  YOUR AD *
*                                                                           *
*  COULOMB: (DIR)=   458.000345    (EXC)=   -21.974411    (TOT)=   436.025934 *
*                                                                           *
*  CONSTR. (MULT)=     0.000000    SLOPE=     0.000000    CORR.=     0.000000 *
*  CONSTR. (SURF)=     0.000000    SLOPE=     0.000000    CORR.=     0.000000 *
*  CONSTR. (SPIN)=    -3.000627    SLOPE=     0.351464    CORR.=    -1.500314 *
*                                                                           *
*  REARRANGEMENT ENERGY FROM THE SKYRME DENSITY-DEPENDENT TERMS=   890.885803 *
*  ROUTHIAN  (TOTAL ENERGY PLUS MULTIPOLE AND SPIN CONSTRAINTS)= -1096.793058 *
*                                                                           *
*  SPIN-ORB (EVE)=   -82.387355    (ODD)=    -0.138618    (TOT)=   -82.525973 *
*  SKYRME:  (EVE)= -3948.363170    (ODD)=    -0.061894    (TOT)= -3948.425064 *
*                                                                           *
*  TOTAL:  (STAB)=     4.76E-07     (SP)= -1093.792430    (FUN)= -1093.792431 *
*                                                                           *
*****************************************************************************
```